\documentclass[sigconf,10pt,authorversion]{acmart}

\usepackage{amsthm}
\theoremstyle{definition}
\newtheorem{example}{Example}[section]

\usepackage{multirow}
\usepackage{color}
\usepackage{graphicx}
\graphicspath{{./figure/}}
\usepackage{balance}
\usepackage{subfig}
\usepackage{hyperref} 
\usepackage{amsmath}
\usepackage{amssymb}
\usepackage[linesnumbered,ruled,vlined]{algorithm2e}

\SetCommentSty{mycommfont}
\usepackage{etoolbox}
\usepackage{adjustbox}

\makeatletter
\newcommand*{\algrule}[1][\algorithmicindent]{%
	\makebox[#1][l]{%
		\hspace*{.2em}
		\vrule height .75\baselineskip depth .25\baselineskip
	}
}
\newcount\ALG@printindent@tempcnta
\def\ALG@printindent{%
	\ifnum \theALG@nested>0
	\ifx\ALG@text\ALG@x@notext
	\else
	\unskip
	\ALG@printindent@tempcnta=1
	\loop
	\algrule[\csname ALG@ind@\the\ALG@printindent@tempcnta\endcsname]%
	\advance \ALG@printindent@tempcnta 1
	\ifnum \ALG@printindent@tempcnta<\numexpr\theALG@nested+1\relax
	\repeat
	\fi
	\fi
}
\newlength{\phaserulewidth}

\makeatother
\usepackage{hyphenat}
\newcommand{\Csharp}{%
	{\settoheight{\dimen0}{C}\kern-.05em \resizebox{!}{\dimen0}{\raisebox{\depth}{\#}}}}
\newcommand{\tony}[1]{{\textcolor{black}{ #1}}}

\newcommand{\tonyy}[1]{{\textcolor{black}{ #1}}}

\newcommand{\system}{BriskStream\xspace}

\newcommand{\compact}{\vspace{-5pt}}
\newcommand{\newcompact}{\vspace{-0pt}}

\begin{document}

\copyrightyear{2019} 
\acmYear{2019} 
\setcopyright{acmcopyright}
\acmConference[SIGMOD '19] {2019 International Conference on Management of Data}{June 30--July 5, 2019}{Amsterdam, Netherlands}
\acmBooktitle{2019 International Conference on Management of Data (SIGMOD '19), June 30--July 5, 2019, Amsterdam, Netherlands}
\acmPrice{15.00}
\acmDOI{10.1145/3299869.3300067}
\acmISBN{978-1-4503-5643-5/19/06} 

\settopmatter{printacmref=false}
\fancyhead{}


\title[BriskStream: Scaling Data Stream Processing on Shared-Memory Multicore Architectures]
{BriskStream: Scaling Data Stream Processing \\ on Shared-Memory Multicore Architectures}

\author{Shuhao Zhang}
\authornote{Work done while as research trainee at SAP Singapore.}
\affiliation{%
  \institution{National University of Singapore}
}

\author{Jiong He}
\affiliation{%
  \institution{Advanced Digital Sciences Center}
}

\author{Amelie Chi Zhou}
\affiliation{%
  \institution{Shenzhen University}
}

\author{Bingsheng He}
\affiliation{%
  \institution{National University of Singapore}
}

\renewcommand{\shortauthors}{S. Zhang, J. He, C. Zhou and B. He}

\begin{abstract}

We introduce \system, an in-memory data stream processing system (DSPSs) specifically designed for modern shared-memory multicore architectures. 
\system's key contribution is an execution plan optimization paradigm, namely RLAS, which takes \emph{relative-location} (i.e., NUMA distance) of \emph{each pair} of producer-consumer operators into consideration. 
We propose a branch and bound based approach with three heuristics to resolve the resulting nontrivial optimization problem. 
The experimental evaluations demonstrate that \system yields much higher throughput and better scalability than existing DSPSs on multi-core architectures when processing different types of workloads.

\end{abstract}  

\maketitle
\section{Introduction}
\label{sec:intro}
Modern multicore processors have demonstrated superior performance for real-world applications~\cite{appuswamy2013scale} with their increasing computing capability and larger memory capacity. 
For example, recent \emph{scale-up} servers can accommodate even hundreds of CPU cores and multi-terabytes of memory~\cite{sgi}. 
\tonyy{Witnessing the emergence of modern commodity machines with massively parallel processors, researchers and practitioners find shared-memory multicore architectures an attractive platform for streaming applications~\cite{profile,StreamBox,SABER}. 
However, prior studies~\cite{profile} have shown that existing data stream processing system (DSPSs) underutilize the underlying complex hardware micro-architecture and show poor scalability due to the unmanaged resource competition and unawareness of non-uniform memory access (NUMA) effect.}


Many DSPSs, such as Storm~\cite{Storm}, Heron~\cite{heron}, Flink~\cite{flink} and Seep~\cite{seep}, share similar architectures including \emph{pipelined} processing and operator \emph{replication} designs.
Specifically, an application is expressed as a DAG (directed acyclic graph) where vertexes correspond to continuously running operators, and edges represent data streams flowing between operators.
To sustain high input stream ingress rates, each operator can be replicated into multiple \emph{replicas} running in parallel threads. 
A \emph{streaming execution plan} determines the number of replicas of each operator (i.e., operator replication), as well as the way of allocating each operator to the underlying CPU cores (i.e., operator placement). 
In this paper, we address the question of how to find a streaming execution plan that maximizes processing throughput of DSPS in shared memory multi-core architectures.


NUMA-aware system optimizations have been previously studied in the context of relational database~\cite{Deployment,Psaroudakis2016,Leis:2014:MPN:2588555.2610507}.
However, those works are either 1) focused on different optimization goals (e.g., better load balancing~\cite{Psaroudakis2016} or minimizing resource utilization~\cite{Deployment}) or 2) based on different system architectures~\cite{Leis:2014:MPN:2588555.2610507}. 
They provide highly valuable techniques, mechanisms and execution models but none of them uses the knowledge at hand to solve the problem we address.

The key challenge of finding an optimal streaming execution plan on multicore architectures is that there is a \emph{varying} processing capability and resource demand of each operator due to \emph{varying} remote memory access penalty under \emph{different} execution plans. 
Witnessing this problem, we present a novel NUMA-aware streaming execution plan optimization paradigm, called \emph{Relative-Location Aware Scheduling} (RLAS).
RLAS takes the relative location  (i.e., NUMA distance) of each pair of producer-consumer into consideration during optimization. 
In this way, it is able to determine the correlation between a solution and its objective value, e.g., predict the throughput of \emph{each} operator for a given execution plan.
This is different to some related studies~\cite{Viglas:2002:RQO:564691.564697, Deployment, Khandekar:2009:COS:1656980.1657002}, which assume a predefined and fixed processing capability (or cost) of each operator.

While RLAS provides a more accurate estimation of the application behavior under the NUMA effect, the resulting placement optimization problem is still challenging to solve. 
In particular, stochasticity is introduced into the problem as the objective value (e.g., throughput) or weight (e.g., resource demand) of each operator is variable and depends on all previous decisions. 
This leads to a huge solution space.
Additionally, the placement decisions may conflict with each other and order constraints are introduced into the problem. 
For instance, scheduling of an operator at one iteration may prohibit some other operators to be scheduled to the same socket later.

We propose a branch and bound based approach to solve the concerned placement optimization problem. 
In order to reduce the size of the solution space, we further introduce three heuristics.
The first switches the placement consideration from vertex to edge, i.e., only consider placement decision of each pair of directly connected operators,
and avoids many placement decisions that have little or no impact on the objective value.
The second reduces the size of the problem in special cases by applying best-fit policy and also avoids identical sub-problems through redundancy elimination. 
The third provides a mechanism to tune the trade-off between optimization granularity and searching space.

RLAS optimizes both replication and placement at the same time. 
The key to optimize replication configuration of a streaming application is to remove bottlenecks in its streaming pipeline.
As each operator's throughput and resource demand may \emph{vary} in different placement plans due to the NUMA effect, removing bottlenecks has to be done together with placement optimization.
To achieve this, RLAS iteratively increases replication level of the bottleneck operator which is identified during placement optimization. 
We implemented RLAS in \system with additional optimizations on shared memory (details in Section~\ref{sec:impl}), a new DSPS supporting the same APIs as Storm and Heron. 
Our extensive experimental study on two eight-socket modern multicores servers show that \system achieves much higher throughput and better scalability than existing DSPSs. 


\textbf{Organization.} 
 The remainder of this paper is organized as follows. 
 Section~\ref{sec:background} covers the necessary background of scale-up servers and an overview of DSPSs. 
 Section~\ref{sec:rlas} discusses the performance model and problem definition of RLAS, followed by a detailed algorithm design in Section~\ref{sec:algo}.
 Section~\ref{sec:impl} {discusses how we optimize \system for shared-memory architectures.} 
 We report extensive experimental results in Section~\ref{sec:eva}. 
 Section~\ref{sec:related} reviews related work and Section~\ref{sec:conclusion} concludes this work.

\compact
\section{Background}
\label{sec:background}
In this section, we introduce modern scale-up servers and give an overview of DSPSs.
\subsection{Modern Scale-up Servers}
Modern machines scale to multiple sockets with non-uniform-memory-access (NUMA) architecture.
Each socket has its own ``local" memory and is connected to other sockets and, hence to their memory, via one or more links.
Therefore, access latency and bandwidth vary depending on whether a core is accessing ``local" or ``remote" memory. Such NUMA effect requires ones to carefully align the communication patterns accordingly to get good performance.
%

\begin{figure}
\begin{minipage}{0.5\textwidth}
    \subfloat[Server A (glue-less) ]{%
        \includegraphics[width=0.5\textwidth]{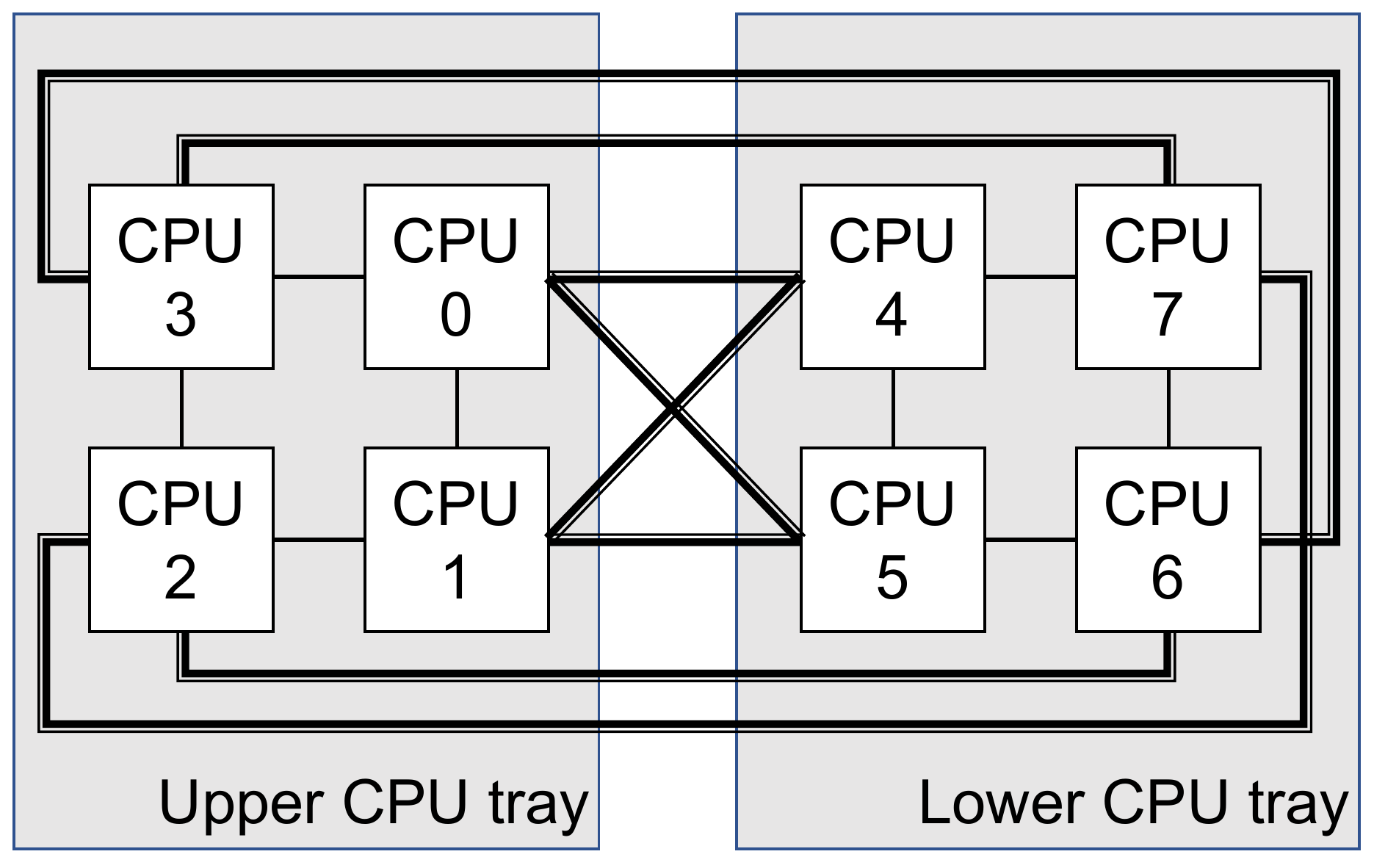}    \label{fig:NUS}
    }  
    \subfloat[Server B (glue-assisted)]{%
        \includegraphics[width=0.5\textwidth]{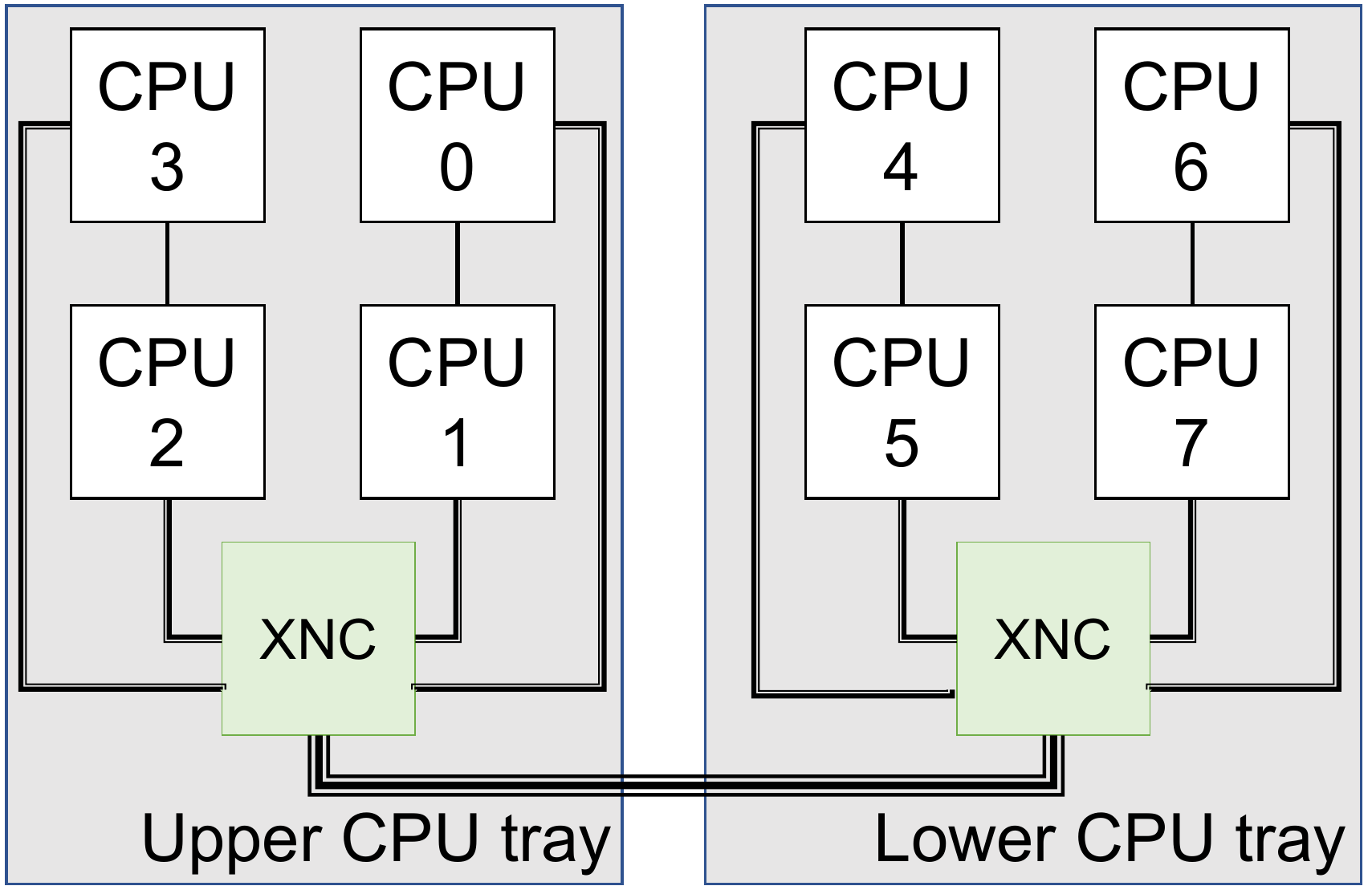}    \label{fig:HPI}    
    }
    \caption{Interconnect topology for our servers.}
    \label{fig:server}
\end{minipage}
\end{figure}

Different NUMA configurations exist in today's market. 
Figure~\ref{fig:server} illustrates the NUMA topologies of our servers in the experiments. 
In the following, we use ``Server A'' to denote the first, and ``Server B'' to denote the second.
Server A can be categorized into the glue-less NUMA server, where CPUs are connected directly/indirectly through QPI or vendor custom data interconnects. 
Server B employs an eXternal Node Controller (called \emph{XNC}~\cite{HPE}) that interconnects upper and lower CPU tray (each tray contains 4 CPU sockets).
The XNC maintains a directory of the contents of each processors cache and significantly reduces remote memory access latency. 
The detailed specifications of our two servers are shown in our experimental setup (Section~\ref{sec:eva}).

\subsection{DSPS Overview}
\begin{figure}
    \subfloat[Logical view of WC.]{%
        \includegraphics[width=0.4\textwidth]{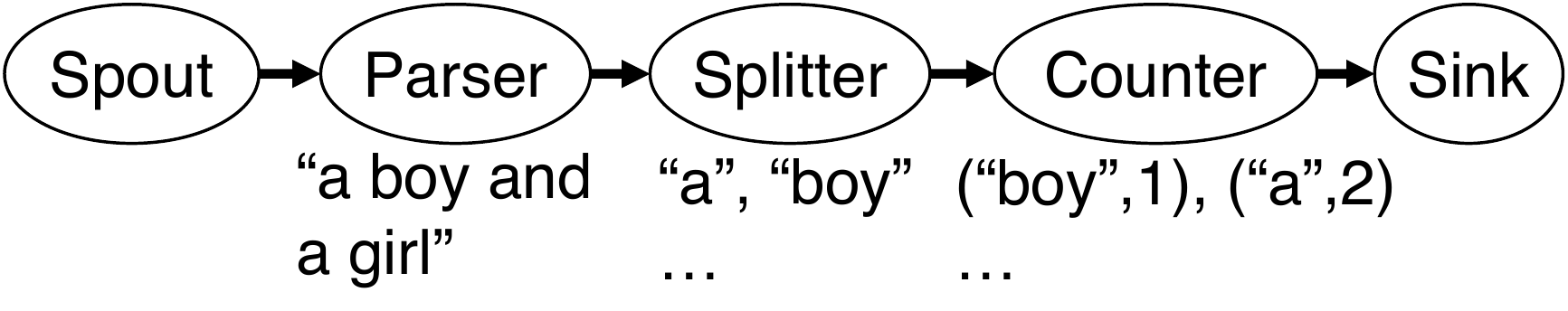}    
    }       
    
    \subfloat[One example execution plan of WC. Three CPU sockets are used.]{%
        \includegraphics[width=0.4\textwidth]{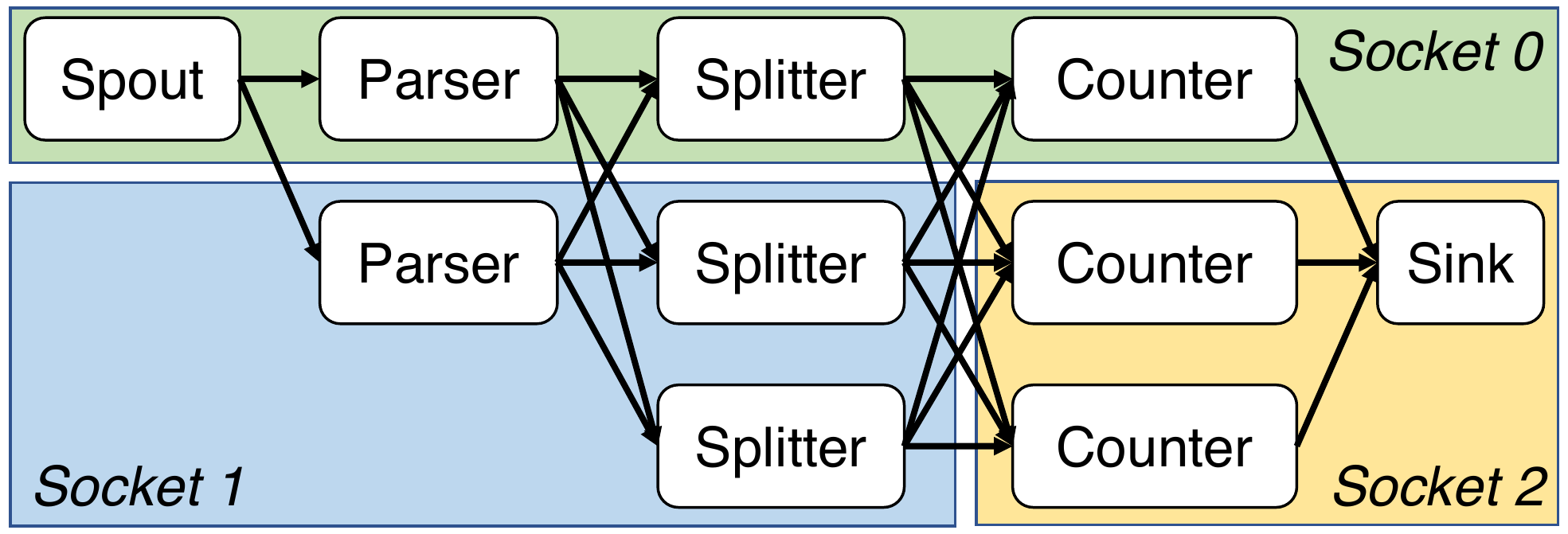}    
    }
    \caption{Word Count (WC) as an example application.}                   
    \label{fig:example}  
\end{figure}

A streaming application is expressed as a DAG (directed acyclic graph) where vertexes correspond to continuously running operators, and edges represent data streams flowing between operators.
Figure~\ref{fig:example}(a) illustrates \emph{word count} (WC) as an example application containing five operators as follows. 
$Spout$ continuously generates new tuple containing a sentence with ten random words. 
$Parser$ drops invalidate tuples (e.g., containing empty value). In our testing workload, the selectivity of the parser is one. 
$Splitter$ processes each tuple by splitting the sentence into words and emits each word as a new tuple to Counter. 
$Counter$ maintains and updates a hashmap with the key as the word and the value as the number of occurrences of the corresponding word. Every time it receives a word from Splitter, it updates the hashmap and emits a tuple containing the word and its current occurrence. 
$Sink$ increments a counter each time it receives tuple from Counter, which we use to monitor the performance of the application.

There are two important aspects of runtime designs of modern DSPSs~\cite{profile}.
First, the common wisdom of designing the execution runtime of DSPSs is to treat each operator as a single execution unit (e.g., a Java thread) and runs multiple operators in a DAG in a pipelining way. 
Second, for scalability, each operator may be executed independently in multiple threads.
Such design is adopted by many DSPSs such as Storm~\cite{Storm}, Flink~\cite{flink}, Seep~\cite{seep}, and Heron~\cite{heron} for its advantage of low processing latency.
Figure~\ref{fig:example}(b) illustrates one example execution plan of WC, where parser, splitter and counter are replicated into 2, 3 and 3 replicas, and they are placed in three CPU sockets (represented as coloured rectangles).

\compact
\section{Execution Plan Optimization}
\label{sec:rlas}
A streaming execution plan concerns how to allocate each operator to underlying physical resources, as well as the number of replicas that each operator should have.
An operator experiences additional remote memory access (RMA) penalty during input data fetch when it is allocated in different CPU sockets to its producers. 
A bad execution plan may introduce unnecessarily high RMA communication overhead and/or oversubscribe a few CPU sockets that induces significant resource contention.
In this section, we discuss the performance model that guides optimization process and the formal definition of our problem.

\subsection{The Performance Model}
\label{sec:model}
Model guided deployment of query plans has been previously studied in relational databases on multi-core architectures, for example~\cite{Deployment}.
Due to the difference in problem assumptions and optimization goals, we adopt a different approach -- the rate-based optimization (RBO) approach~\cite{Viglas:2002:RQO:564691.564697}, where output rate of each operator is estimated.
However, the original RBO~\cite{Viglas:2002:RQO:564691.564697} assumes processing capability of an operator is predefined and independent of execution plans, which is not suitable under the NUMA effect.

We summarize the main terminologies of our performance model in Table~\ref{tbl:notations}. 
We group them into the following four types, including \emph{machine specifications}, \emph{operator specifications},  \emph{plan inputs} and \emph{model outputs}. For the sake of simplicity, we refer a replica of an operator simply as an ``operator''. 
Machine specifications are the information of the underlying hardware.
Operator specifications are the information specific to an operator, which need to be directly profiled (e.g., $T^{e}$) or indirectly estimated with profiled information and model inputs (e.g., $T^{f}$).
Plan inputs are the specification of the execution plan including both placement and replication plans as well as external input rate to the source operator.
Model outputs are the final results of the performance model. 
To simplify the presentation, we omit the selectivity estimation and assume selectivity is one in the following discussion. 
In our experiment, the selectivity statistics of each operator are pre-profiled before the optimization applies.
In practice, they can be periodically collected during runtime and the optimization needs to be re-performed accordingly.

\begin{table}
  \centering
  \caption{Summary of terminologies}
  \label{tbl:notations}
  \includegraphics[width=0.45\textwidth]{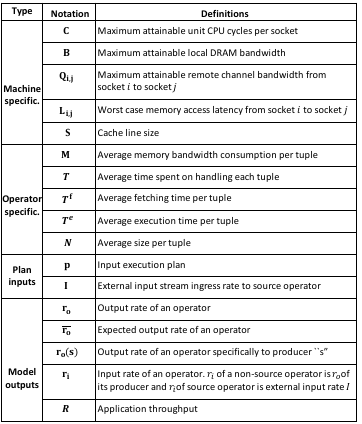}
\end{table}

\textbf{Model overview.}
In the following, we refer to the output rate of an operator using the symbol $r_o$, while $r_i$ refers to its input rate. 
%
The throughput ($R$) of the application is modelled as the summation of $r_o$ of all sink operators (i.e., operators with no consumer). That is, $R=\sum_{sink}{r_o}.$ To estimate R, we hence need to estimate $r_o$ of each sink operator. The output rate of an operator is not only related to its input rate but also the execution plan due to NUMA effect, which is quite different from previous studies~\cite{Viglas:2002:RQO:564691.564697}. 

As \system adopted the pass-by-reference message passing approach (See Appendix~\ref{sec:impl_detail}) to utilize shared-memory environment, the reference passing delay is negligible. 
Hence, $r_i$ of an operator is simply $r_o$ of the corresponding producer and $r_i$ of spout (i.e., source operator) is given as $I$ (i.e., external input stream ingress rate). 
Conversely, upon obtaining the reference, an operator then needs to fetch the actual data during its processing, where the actual data fetch delay depends on NUMA distance between it and its producer. 
We hence estimate $r_o$ of an operator as a function of its input rate $r_i$ and execution plan $p$.


\textbf{Estimating $r_o$.}
Consider a time interval $t$, denote the number of tuples to be processed during $t$ as $num$ and actual time needed to process them as $t_p$. 
Further, denote $T(p)$ as the average time spent on handling each tuple for a given execution plan $p$.
Let us first assume input rate to the operator is sufficiently large and the operator is always busy during $t$ (i.e., $t_p > t$), and we discuss the case of $t_p \leq t$ at the end of this paragraph. Then, the general formula of $r_o$ can be expressed in Formula~\ref{eqn:ro}. 
Specifically, $num$ is the total number of input tuples from all producers arrived during $t$, and $t_p$ is the total time spent on processing those input tuples. 

\begin{align}
   r_o & = \frac{num}{t_p},  \nonumber\\
   where\ num &=\sum_{producers} r_i\times t\nonumber\\ 
          t_p &=\sum_{producers} {r_i\times t}\times{T(p)}.
   \label{eqn:ro}
\end{align}

We breakdown $T(p)$ into the following two non-overlapping components, $T^{e}$ and $T^{f}$ (i.e., $ T(p)=T^{e}+T^{f}$).

$T^{e}$ stands for time required in actual function execution and emitting output tuples per input tuple. 
For operators that have a constant workload for each input tuple, we simply measure its average execution time per tuple with one execution plan to obtain its $T^e$. Otherwise, we can use machine learning techniques (e.g., linear regression) to train a prediction model to predict its $T^e$ under varying execution plans. Prediction of an operator with more complex behaviour has been studied in previous works~\cite{6228100}, and we leave it as future work to enhance our system.

$T^{f}$ stands for time required to (locally or remotely) fetch the actual data per input tuple. It is determined by its fetched tuple size and its relative distance to its producer (determined by $p$), which can be represented as follows,

\begin{align}
    T^{f}&=
     \begin{cases}
       0       & \text{if collocated with producer}  \\
       \lceil N/S\rceil\times L_{(i,j)}       & \text{otherwise}
      \end{cases}\nonumber\\
      & \text{ where $i$ and $j$ are determined by }p.
    \label{eqn:fecycles}
\end{align}

When the operator is collocated with its producer, the data fetch cost is already covered by $T^{e}$ and hence $T^{f}$ is 0.
Otherwise, it experiences memory access across CPU sockets per tuple.
It is generally difficult to accurately estimate the actual data transfer cost as it is affected by multiple factors such as memory access patterns and hardware prefetcher units.
We use a simple formula based on a prior work~\cite{byna2004predicting} as illustrated in Formula~\ref{eqn:fecycles}. 
Specifically, we estimate the cross socket communication cost based on the total size of data transfer $N$ bytes per input tuple, cache line size $S$ and the worst case memory access latency ($L_{(i,j)}$) that operator and its producer allocated ($i\neq j$). 
Applications in our testing benchmark roughly follow Formula~\ref{eqn:fecycles} as we show in our experiments later.

Finally, let us remove the assumption that input rate to an operator is larger than its capacity, and denote the expected output rate as $\overline{r_o}$. There are two cases that we have to consider:

\begin{itemize}
\label{itemize}
    \item[Case 1:] We have essentially made an assumption that the operator is in general \emph{over-supplied}, i.e., $t_p \ge t$. In this case, input tuples are accumulated and $\overline{r_o}=r_o$. As tuples from all producers are processed in a cooperative manner with equal priority, tuples will be processed in a first come first serve manner. It is possible to configure different priorities among different operators here, which is out of the scope of this paper. 
    Therefore, $r_o(s)$ is determined by the proportion of the corresponding input ($r_i(s)$), that is, $\overline{r_o(s)}=r_o\times\frac{r_i(s)}{r_i}$.
    \item[Case 2:] In contrast, an operator may need less time to finish processing all tuples arrived during observation time $t$, i.e., $t_p < t$. In this case, we can derive that $r_o \geq \sum_{producers} r_i$. This effectively means the operator is \emph{under-supplied}, and its output rate is limited by its input rates, i.e., $\overline{r_o} = r_i$, and $\overline{r_o(s)} = r_i(s)\ \forall\ \text{producer }s$. 
\end{itemize}

\begin{figure}
    \centering
    \includegraphics*[width=0.4\textwidth]{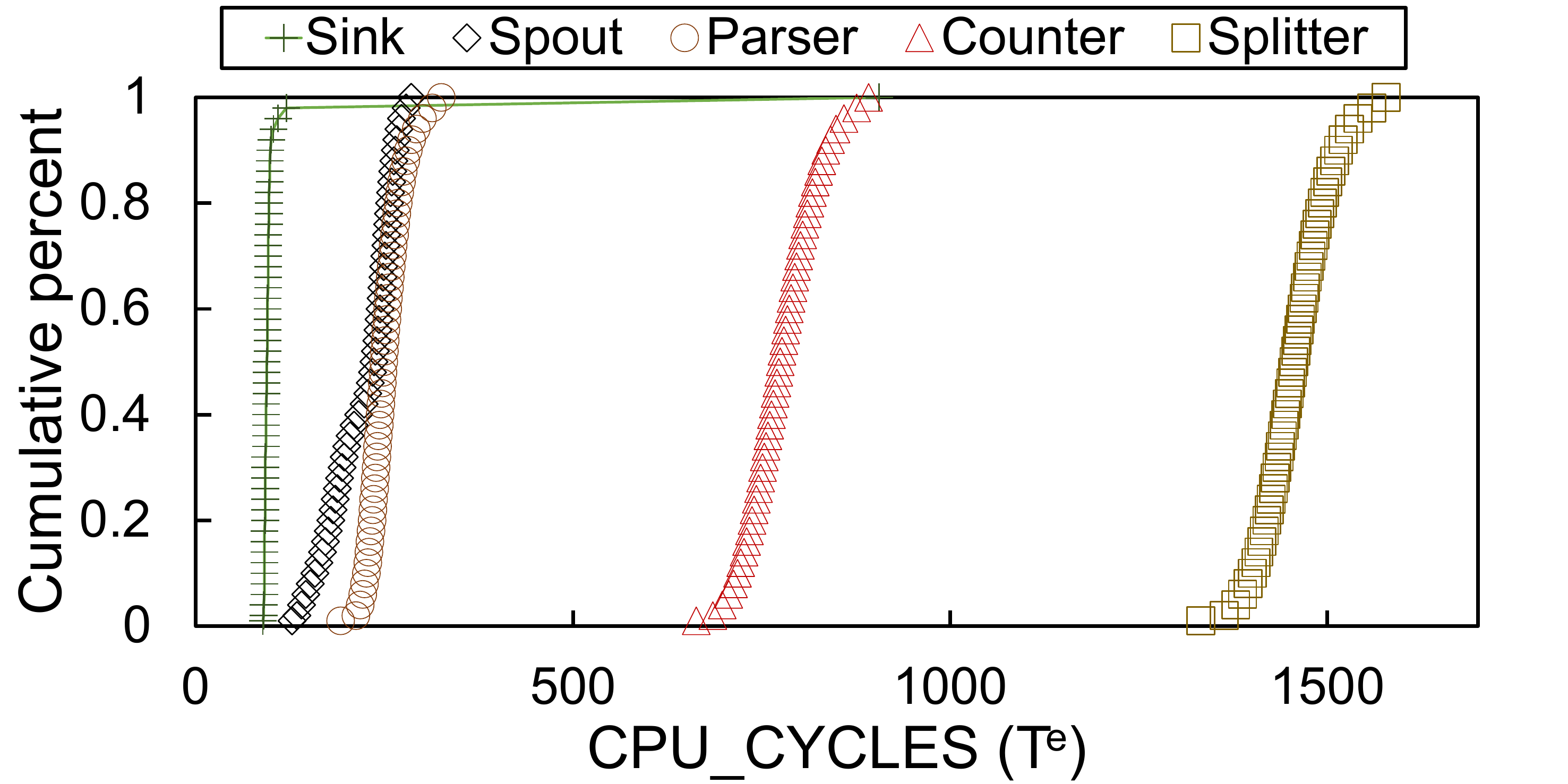}
    \caption{CDF of profiled average execution cycles of different operators of WC.}  \label{fig:profile_distribution}   
\end{figure}

Given an execution plan, we can then identify operators that are over-supplied by comparing its input rate and output rate. 
Those over-supplied operators are essentially the ``bottlenecks'' of the corresponding execution plan. Our scaling algorithm tries to increase the replication level of those operators to remove bottlenecks. After the scaling, we need to again look for the optimal placement plan of the new DAG. This iterative optimization process formed our optimization framework, which will be discussed shortly later in Section~\ref{sec:algo}.

\textbf{Model instantiation.}
Machine specifications of the model including $C$, $B$, $Q_{i,j}$, $L_{i,j}$ and $S$ are given as statistics information of the targeting machine (e.g., measured by Intel Memory Latency Checker~\cite{intel}). 
Similar to the previous work~\cite{Cheung2013SpeedingUD}, we need to profile the application to determine operator specifications.
To eliminate the impact of interference, we sequentially profile each operator. 
Specifically, we first launch a \emph{profiling thread} of the operator to profile on one core. 
Then, we feed sample input tuples (stored in local memory) to it. 
Information including $T^{e}$ (execution time per tuple), $M$ (average memory bandwidth consumption per tuple) and $N$ (size of input tuple) is then gathered during its execution.

The sample input is prepared by pre-executing all upstream operators. As they are not running during profiling, they will not interfere with the profiling thread. 
To speed up the instantiation process, multiple operators can be profiled at the same time as long as there is no interference among the profiling threads (e.g., launch them on different CPU sockets). 
The statistics gathered without interference are used in the model as \system avoids interference (see Section~\ref{subsec:def}). Task oversubscribing has been studied in some earlier work~\cite{5470434}, but it is not the focus of this paper.

We use the overseer library~\cite{overseer} to measure $T^{e}$, $M$, and use classmexer library~\cite{classmexer} to measure $N$.
Figure~\ref{fig:profile_distribution} shows the profiling results of $T^{e}$ of different operators of WC. 
The major takeaway from Figure~\ref{fig:profile_distribution} is that operators show stable behaviour in general, and the statistics can be used as model input. Selecting a lower (resp. higher) percentile profiled results essentially corresponds to a more (resp. less) optimistic performance estimation. Nevertheless, we use the profiled statistics at the \emph{50th} percentile as the input of the model, which sufficiently guides the optimization process.


\subsection{Problem Formulation}
\label{subsec:def}
The goal of our optimization is to maximize the application processing throughput under given input stream ingress rate, where we look for the optimal replication level and placement of each operator. 
For one CPU socket, denote its available CPU cycles as $C$ cycles/sec, the maximum attainable local DRAM bandwidth as $B$ bytes/sec, and the maximum attainable remote channel bandwidth from socket $S_i$ to $S_j$ as $Q_{i,j}$ bytes/sec. 
Further, denote average tuple size, memory bandwidth consumption and processing time spent per tuple of an operator as $N$ bytes, $M$ bytes/sec and $T$ cycles, respectively, 
The problem can be mathematically formulated as Equation~\ref{eqn:amountC}--\ref{eqn:constraint_q}.

As the formulas show, we consider three categories of resource constraints that the optimization algorithm needs to make sure the execution plan satisfies. 
Constraint in Eq.~\ref{eqn:amountC} enforces that the aggregated demand of CPU resource requested to anyone CPU socket must be smaller than the available CPU resource.
Constraint in Eq.~\ref{eqn:constraint_m} enforces that the aggregated amount of bandwidth requested to a CPU socket must be smaller than the maximum attainable local DRAM bandwidth. Constraint in Eq.~\ref{eqn:constraint_q}
enforces that the aggregated data transfer from one socket to another per unit of time must be smaller than the corresponding maximum attainable remote channel bandwidth.
In addition, it is also constrained that one operator is allocated exactly once. This matters because an operator may have multiple producers that are allocated at different places. In this case, the operator can only be collocated with a subset of its producers. 

\begin{align}
    &maximize \sum_{sink} \overline{r_o}\nonumber\\
    &\text{s.t., $\forall i,j \in {1,..,n}$,} \nonumber\\
    &\sum_{operators\ at\ S_i}\overline{r_o}*T\leq C, \label{eqn:amountC}    \\
    &\sum_{operators\ at\ S_i}\overline{r_o}*M\leq B,
            \label{eqn:constraint_m}\\
    &\sum_{operators\ at\ S_j}\sum_{producers\ at\ S_i}\overline{r_o(s)}*N\leq Q_{i,j},     \label{eqn:constraint_q}
\end{align}

Assuming each operator (suppose in total $|o|$ operators) can be replicated at most $k$ replicas, we have to consider in total $k^{|o|}$ different replication configurations.
In addition, for each replication configuration, there are $m^{n}$ different placements, where $m$ is the number of CPU sockets and $n$ stands for the total number of replicas ($n\geq |o|$). 
Such a large solution space makes brute-force unpractical.

\section{Optimization Algorithm Design} 
\label{sec:algo}
We propose a novel optimization paradigm called \emph{Relative-Location Aware Scheduling} (RLAS) to optimize replication level and operator placement at the same time guided by our performance model. 
The key to optimize replication configuration of a stream application is to remove bottlenecks in its streaming pipeline.
As each operator's throughput and resource demand may \emph{vary} in different placement plans, removing bottlenecks has to be done together with placement optimization.

\begin{figure}
    \centering
    \includegraphics*[width=0.48\textwidth]{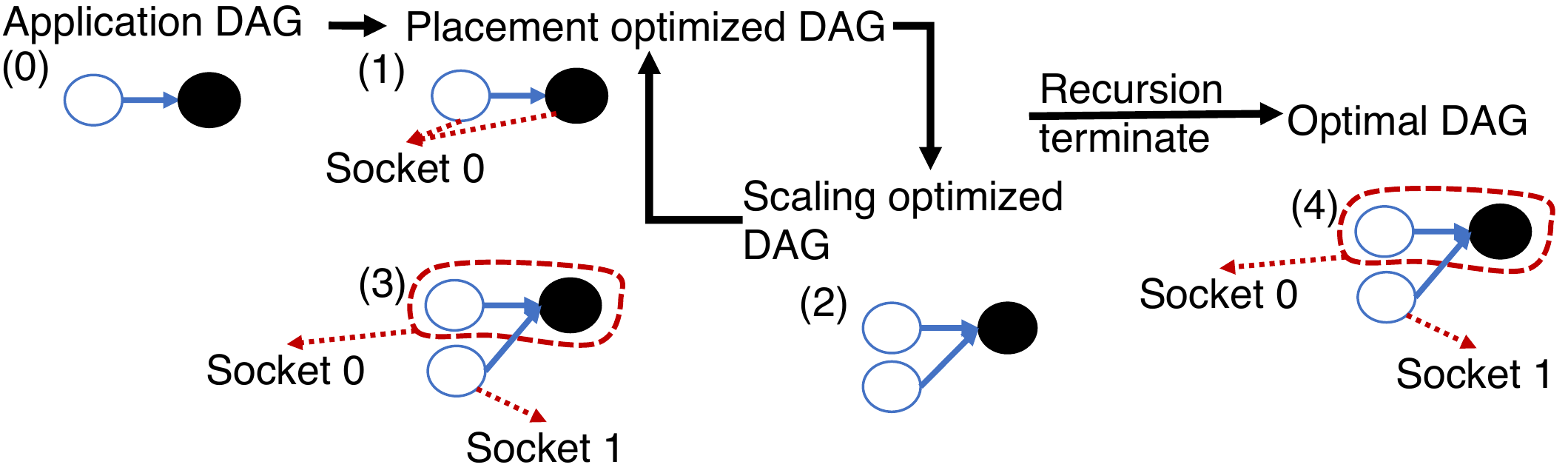}
    \caption{RLAS Optimization example.}  \label{fig:BriskStream} 
\end{figure}

The key idea of our optimization process is to iteratively optimize operator placement under a given replication level setting and then try to increase replication level of the bottleneck operator, which is determined during placement optimization. 
Specifically, the operator that is overfed is defined as bottleneck (see Case 1 in Section~\ref{itemize}).
Figure~\ref{fig:BriskStream} shows an optimization example of a simple application consisting of two operators. 
The initial execution plan with no operator replication is labelled with (0).
First, RLAS optimizes its placement (labelled with (1)) with \emph{placement algorithm}, which also identifies bottleneck operators. 
The operators' placement to CPU sockets are indicated by the dotted arrows in the Figure.
Subsequently, it tries to increase the replication level of the bottleneck operator, i.e., the hollow circle, with \emph{scaling algorithm} (labelled with (2)). 
It continues to optimize its placement given the new replication level setting (labelled with (3)).
Finally, the application with an optimized execution plan (labelled with (4)) is submitted to execute.



The details of scaling and placement optimization algorithms are presented in Appendix~\ref{subsec:algo_impl}. In the following, we discuss how the Branch and Bound (B\&B) based technique~\cite{MORRISON201679} is applied to solve our placement optimization problem assuming operator replication is given as input. 
We focus on discussing our bounding function and proposed heuristics that improve the searching efficiency. 

\textbf{Branch and Bound Overview.} 
B\&B systematically enumerates a tree with nodes representing candidate solutions, based on a bounding function. 
There are two types of nodes in the tree: live nodes and solution nodes. 
In our context, a node represents a placement plan and the value of a node stands for the estimated throughput under the corresponding placement.
A \emph{live node} contains the placement plan that violates some constraints and they can be expanded into other nodes that violate fewer constraints. 
The value of a live node is obtained by evaluating the bounding function.
A \emph{solution node} contains a valid placement plan without violating any constraint. 
The value of a solution node comes directly from the performance model. 
The algorithm may reach multiple solution nodes as it explores the solution space. 
The solution node with the best value is the output of the algorithm. 

\emph{Algorithm complexity:} 
Naively in each iteration, there are  $\binom{n}{1}*\binom{m}{1}=n*m$ possible solutions to branch, i.e., schedule \emph{which} operator to \emph{which} socket and an average $n$ depth as one operator is allocated in each iteration. 
In other words, it will still need to examine on \emph{average} $(n*m)^{n}$ candidate solutions~\cite{Devroye:1999:CBB:329857.329859}.
In order to further reduce the complexity of the problem, heuristics have to be applied.

\textbf{The bounding function.} Specifically, the bounded value of every live node is obtained by fixing the placement of \emph{valid} operators and let \emph{remaining} operators to be collocated with all of its producers, which may violate resource constraints as discussed before, but gives the upper bound of the output rate that the current node can achieve.
If the bounding function value of an intermediate node is worse than the solution node obtained so far, we can safely prune it and all of its children nodes. 
This does not affect the optimality of the algorithm because the value of a live node must be better than all its children node after further exploration. 
In other words, the value of a live node is the theoretical upper bound of the subtree of nodes.
The bounded problem that we used in our optimizer originates from the same optimization problem with relaxed constraints.

Consider a simple application with operators A, A' (replica of A) and B, where A and A' are producers of B.
Assume at one iteration, A and A' are scheduled to socket 0 and 1, respectively (i.e., they become valid). 
We want to calculate the bounding function value assuming B is the sink operator, which remains to be scheduled.
In order to calculate the bounding function value, we simply let B be collocated with both A and A' at the same time, which may violate some constraints. In this way, its output rate is maximized, which is the bounding value of the live node.
The calculating of our bounding function has the same cost as evaluating the performance model since we only need to mark $T^{f}$ (Formula~\ref{eqn:fecycles}) to be $0$ for those operators remaining to be scheduled. 

\textbf{The branching heuristics.} 
We introduce the following three heuristics that work together to significantly reduce the solution space.

{\emph{1) Collocation heuristic}:}
The first heuristic switches the placement consideration from vertex to edge, i.e., only consider placement decision of each pair of directly connected operators.
This avoids many placement decisions of a single operator that have little or no impact on the output rate of other operators.
Specifically, the algorithm considers a list of \emph{collocation} decisions involving a pair of directly connected producer and consumer. 
During the searching process, collocation decisions are gradually removed from the list once they become no longer relevant. 
For instance, it can be safely discarded (i.e., do not need to consider anymore) if both producer and consumer in the collocation decision are already allocated.
  
{\emph{2) Best-fit \& Redundant-elimination heuristic}:}
The second reduces the size of the problem in special cases by applying best-fit policy and also avoids identical sub-problems through redundancy elimination. 
Consider an operator to be scheduled, if all predecessors (i.e., upstream operators) of it are already scheduled, then the output rate of it can be safely determined without affecting any of its predecessors.
In this case, we select only the best way to schedule it to maximize its output rate. 
Furthermore, in case that there are multiple sockets that it can achieve maximum output rate, we only consider the socket with the least remaining resource. If there are multiple equal choices, we only branch to one of them to reduce problem size.
 
{\emph{3) Compress graph}:}
The third provides a mechanism to tune a trade-off between optimization granularity and searching space.
Under a large replication level setting, the execution graph becomes very large and the searching space is huge. 
We compress the execution graph by grouping multiple replicas of an operator (denoted by \emph{compress ratio}) into a single large instance that is scheduled together. 
Essentially, the compress ratio represents the tradeoff between the optimization granularity and searching space. 
By setting the ratio to be one, we have the most fine-grained optimization but it takes more time to solve. 
In our experiment, we set the ratio to be 5, which produces a good trade-off.

\begin{figure}
    \centering
    \includegraphics[width=.48\textwidth]{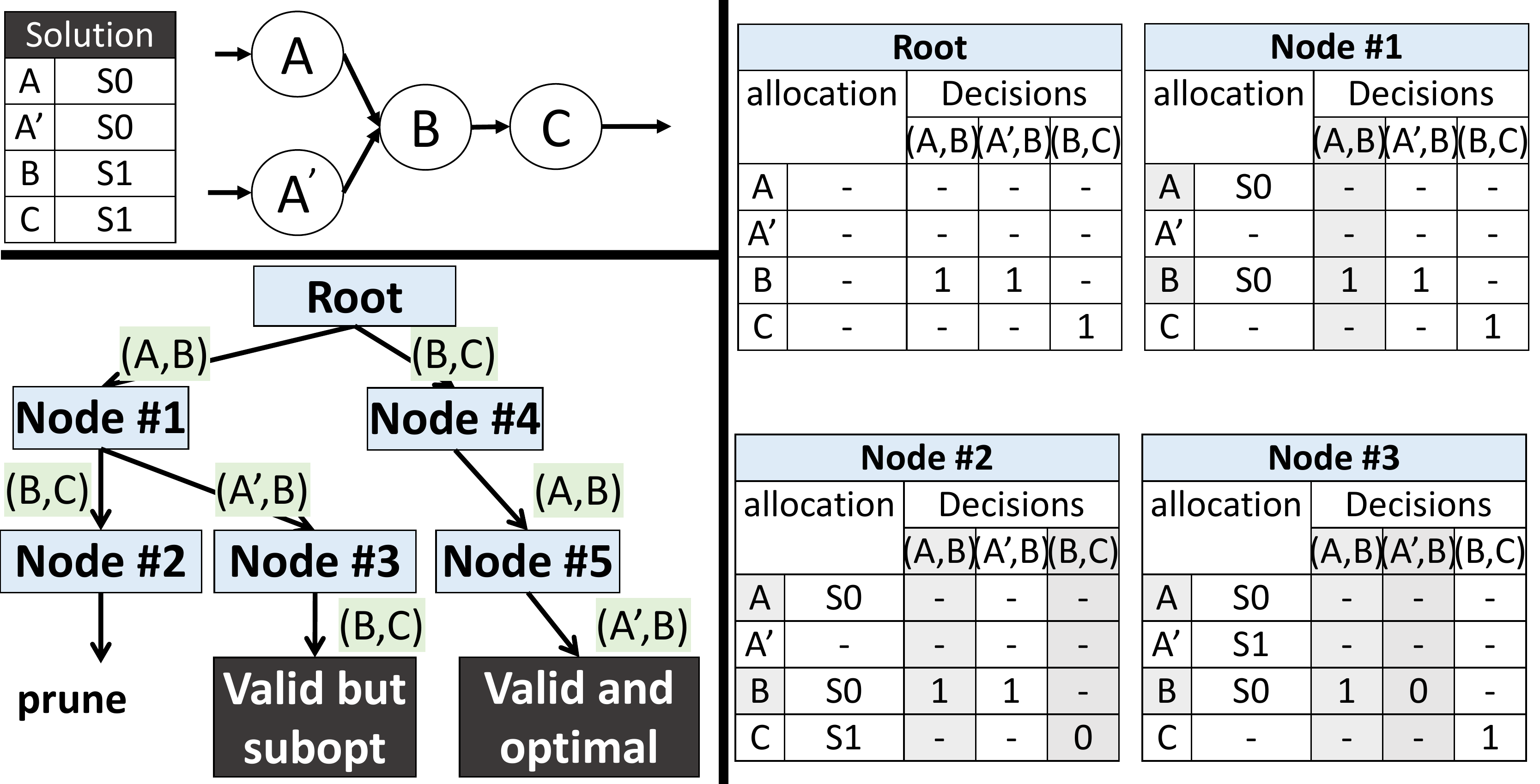}
    \caption{Placement optimization at runtime. Light colored rectangle represents a live node that still violates resource constraints. Dark colored rectangle stands for a solution node contains a valid plan.}
    \label{fig:algorithm}
\end{figure}

We use the scheduling of WC as a concrete example to illustrate the algorithm. 
For the sake of simplicity, we consider only an intermediate iteration of scheduling of a subset of WC. 
Specifically, two replicas of the parser (denoted as $A$ and $A'$), one replica of the splitter (denoted as $B$), and one replica of count (denoted as $C$) are remaining unscheduled as shown in the top-left of Figure~\ref{fig:algorithm}.

In this example, we assume the aggregated resource demands of any combinations of grouping three operators together exceed the resource constraint of a socket, and the only optimal scheduling plan is shown beside the topology.
The bottom left of the Figure shows how our algorithm explores the searching space by expanding nodes, where the label on the edge represents the collocation decision considered in the current iteration.
The detailed states of four nodes are illustrated on the right-hand side of the figure, where the state of each node is represented by a two-dimensional matrix. 
The first (horizontal) dimension describes a list of collocation decisions, 
while the second one represents the operator that interests in this decision. 
A value of `-' means that the respective operator is not interested in this collocation decision.
A value of `1' means that the collocation decision is made in this node, although it may violate resource constraints. An operator is interested in the collocation decision involving itself to minimize its remote memory access penalty.
A value of `0' means that the collocation decision is not satisfied and the involved producer and consumer are separately located. 

At the root node, we consider a list of scheduling decisions involving each pair of producer and consumer. 
At Node \Csharp{}1, the collocation decision of A and B is going to be satisfied, and assume they are collocated to S0. 
Note that, S1 is identical to S0 at this point and does not need to repeatedly consider.
The bounding value of this node is essentially collocating all operators into the same socket, and it is larger than solution node hence we need to further explore. 
At Node \Csharp{}2, we try to collocate A' and B, which however cannot be satisfied (due to the assumed resource constraint). 
As its bounding value is worse than the solution (if obtained), it can be pruned safely.
Node \Csharp{}3 will eventually lead to a valid yet bad placement plan. 
One of the searching processes that leads to the solution node is Root$\rightarrow$Node \Csharp{}4$\rightarrow$Node\Csharp{}5$\rightarrow$Solution.

\compact
\vspace{-1pt}
\section{BriskStream System}
\label{sec:impl}

Applying RLAS to existing DSPSs (e.g., Storm, Flink, Heron) is insufficient to make them scale on shared-memory multicore architectures. 
As they are not designed for multicore environment~\cite{profile}
, much of the overhead come from the inherent distributed system designs. 


We integrate RLAS optimization framework into \system~\footnote{The source code of \system will be publicly available at~\url{https://github.com/ShuhaoZhangTony/briskstream}.}, a new DSPS supporting the same APIs as Storm and Heron. 
More implementation details of \system are given in Appendix~\ref{sec:impl_detail}. 
According to Equation~\ref{eqn:ro}, both $T^{e}$ and $T^{f}$ shall be reduced in order to improve output rate of an operator and subsequently improve application throughput. 
In the following, we discuss two design aspects of \system that are specifically optimized for shared-memory architectures that reduce $T^{e}$ and $T^{f}$ significantly. 
We also discuss some limitations in Section~\ref{sec:limits}.

\subsection{Improving Execution Efficiency}
\label{subsec:impl1}
Compared with distributed DSPSs, BrickStream eliminates many unnecessary components to reduce the instruction footprint, notably including (de)serialization, cross-process and network-related communication mechanism, and condition checking (e.g., exception handling). 
Those unnecessary components (although not involved during execution) bring many conditional branch instructions and results in large instruction footprint~\cite{profile}.
Furthermore, we carefully revise the critical execution path to avoid unnecessary/duplicate temporary object creations. 
For example, as an output tuple is exclusively accessible by its targeted consumer and all operators share the same memory address, we do not need to create a new instance of the tuple when the consumer obtains it. 

\subsection{Improving Communication Efficiency}
\label{subsec:impl2}
Most modern DSPSs~\cite{Storm,flink, profile} employ buffering strategy to accumulate multiple tuples and send them in batches to improve the application throughput. 
\system follows the similar idea of buffering output tuples, but accumulated tuples are combined into one ``jumbo tuple'' (see the example in Appendix~\ref{sec:impl_detail}). 
This approach has several benefits for scalability. 
First, since we know tuples in the same jumbo tuple are targeting at the same consumer from the same producer in the same process, we can eliminate duplicate tuple header (e.g., metadata, context information) hence reduces communication costs.
In addition, the insertion of a jumbo tuple (containing multiple output tuple) requires only a single insertion to the communication queue and effectively amortizing the insertion overhead. 
As a result, both $T^{e}$ and $T^{f}$ are significantly reduced.


\subsection{Discussions}
\label{sec:limits}
\tony{To examine the maximum system capacity, we assume input stream ingress rate ($I$) is sufficiently large and keeps the system busy. 
Hence, the model instantiation and subsequent execution plan optimization are conducted at the same \emph{over-supplied} configuration. 
In practical scenarios, stream rate as well as its characteristics can vary over time, and application needs to be re-optimized in response to workload changes~\cite{7568388,Elastic,Schneider2016}. 
To adapt our optimizations to dynamic scenarios, we plan to study simple heuristic algorithms such as round-robin or traffic-minimization allocation~\cite{profile,T-storm}.
}

\compact
\section{Evaluation}
\label{sec:eva}
Our experiments are conducted in following aspects.
First, our proposed performance model accurately predict the application throughput under different execution plans (Section~\ref{subsec:model}).
Second, BriskStream significantly outperforms existing open-sourced DSPSs on multicores (Section~\ref{subsec:design}).
Third, our RLAS optimization approach performs significantly better than competing techniques (Section~\ref{subsec:rlas}).
We also show in Section~{\ref{subsec:factor}} the relative importance of BriskStream's optimization techniques. 

\subsection{Experimental Setup}
\label{subsec:setup}
We pick four common applications from the previous study~\cite{profile} with different characteristics to evaluate BriskStream.
These tasks are word-count (WC), fraud-detection (FD), spike-detection (SD), and linear-road (LR) with different topology complexity and varying compute and memory bandwidth demand. 
More application settings can be found in Appendix~\ref{subsec:application}. 

To examine the maximum system capacity under given hardware resources, we tune the input stream ingress rate ($I$) to its maximum attainable value ($I_{max}$) to keep the system busy and report the stable system performance~\footnote{Back-pressure mechanism will eventually slow down spout so that the system is stably running at its best achievable throughput.}. 
To minimize interference of operators, we use OpenHFT Thread Affinity Library~\cite{openhft} with core isolation (i.e., configure $isolcpus$ to avoid the isolated cores being used by Linux kernel general scheduler) to bind operators to cores based on the given execution plan. 

\begin{figure}   
    \centering
        \captionof{table}{Characteristics of the two servers we use}
        \label{tbl:NUMAservers}
    \includegraphics[width=0.4\textwidth]{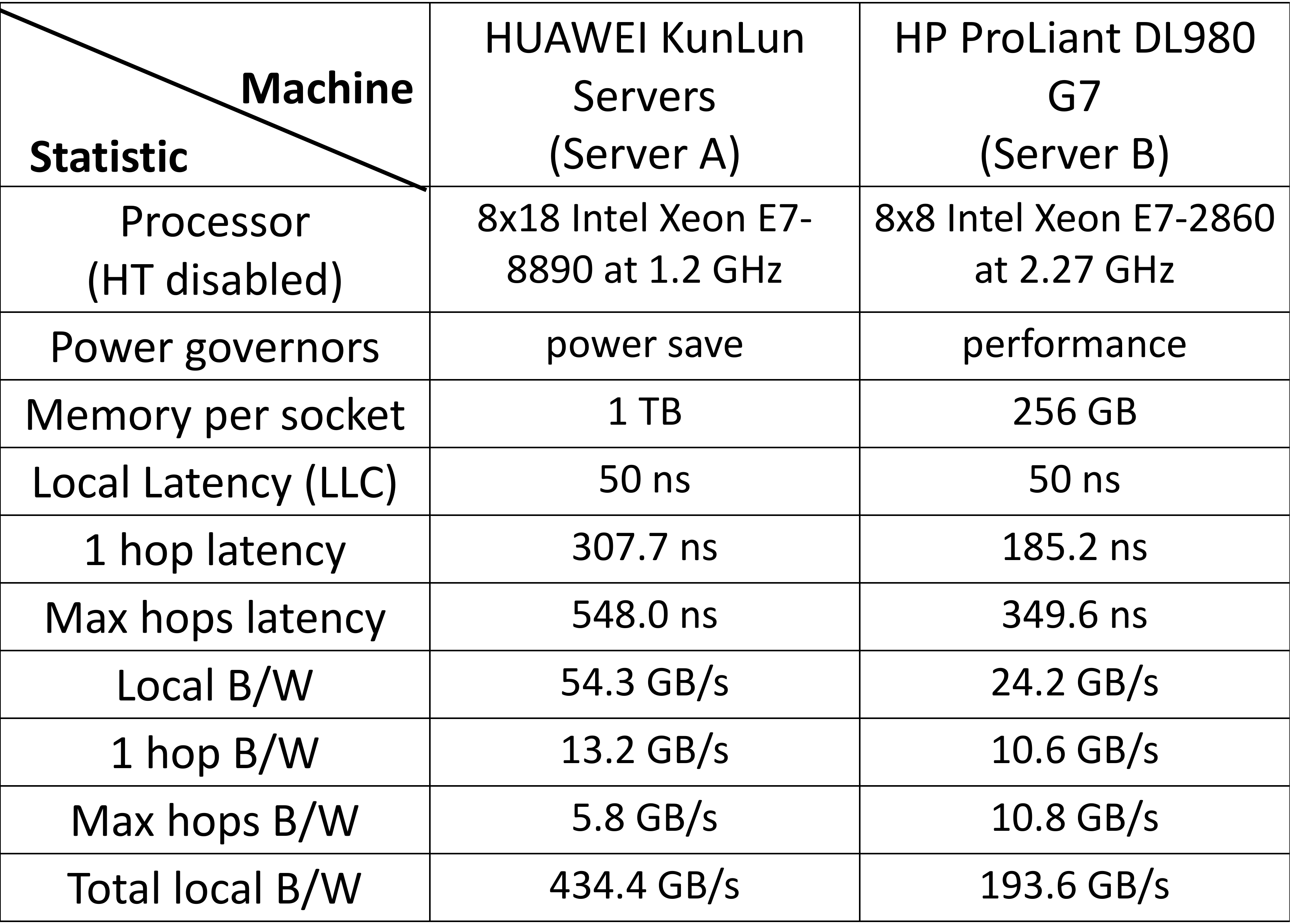} 
\end{figure}

Table~\ref{tbl:NUMAservers} shows the detailed specification of our two eight-socket servers. 
We use Server A in Section~\ref{subsec:model},~\ref{subsec:design} and~\ref{subsec:factor}. 
We study our RLAS optimization algorithms in detail on different NUMA architectures with both two servers in Section~\ref{subsec:rlas}. 
NUMA characteristics, such as local and inter-socket idle latencies and peak memory bandwidths, are measured with Intel Memory Latency Checker~\cite{intel}.
These two machines have different NUMA topologies, which lead to different access latencies and throughputs across CPU sockets. 
The three major takeaways from Table~\ref{tbl:NUMAservers} are as follows. 
First, due to NUMA, both Servers have significantly high remote memory access latency, which is up to 10 times higher than local cache access.
Second, different interconnect and NUMA topologies lead to quite different bandwidth characteristics on these two servers. 
In particular, remote memory access bandwidth is similar regardless of the NUMA distance in Server B. 
In contrast, the bandwidth is significantly lower across long NUMA distance than smaller distance on Server A.
Third, there is a significant increase in remote memory access latency from within the same CPU tray (e.g., 1 hop latency) to between different CPU trays (max hops latency) on both servers. 

In addition to runtime statistics evaluation, we also report how much time each tuple spends in different components of the system. 
We classify these components as follows:
\emph{1) Execute} refers to the average time spent in core function execution. 
Besides the actual user function execution, it also includes various processor stalls such as instruction cache miss stalls.
\emph{2) RMA} refers to the time spend due to remote memory access. This is only involved when the operator is scheduled to different sockets to its producers, and it varies depending on the relative location between operators.
\emph{3) Others} consist of all other time spent in the critical execution path and considered as overhead. 
Examples include temporary object creation, exception condition checking, communication queue accessing and context switching overhead.

To measure \emph{Execute} and \emph{Others}, we allocate the operator to be collocated with its producer.
The time spend in user function per tuple is then measured as \emph{Execute}.
We measure the gap between the subsequent call of the function as \emph{round-trip delay}.
\emph{Others} is then derived as the subtraction from \emph{round-trip delay} by \emph{Execute}.
Note that, the measurement only consists of contiguous successful execution and exclude the time spend in queue blocking (e.g., the queue is empty or full).  
To measure \emph{RMA} cost, we allocate the operator remotely to its producer and measures the new \emph{round-trip delay} under such configuration. 
The \emph{RMA} cost is then derived as the subtraction from the new \emph{round-trip delay} by the original \emph{round-trip delay}.

\subsection{Performance Model Evaluation}
\label{subsec:model}
In this section, we evaluate the accuracy of our performance model.
We first evaluate the estimation of the cost of remote memory access. 
We take Split and Count operators of WC as an example. 
Table~\ref{tbl:sub_accuracy} compares the measured and estimated process time per tuple ($T$) of each operator.
Our estimation generally captures the correlations between remote memory access penalty and NUMA distance. 
The estimation is larger than measurement, especially for Splitter. 
When the input tuple size is large (in case of Splitter), the memory accesses have better locality and the hardware prefetcher helps in reducing communication cost~\cite{Lee:2012:PWD:2133382.2133384}.
Another observation is that there is a significant increase of RMA cost from between sockets from the same CPU tray (e.g., S0 to S1) to between sockets from different CPU tray (e.g., S0 to S4). Such non-linear increasing of RMA cost has a major impact on the system scalability as we need to pay significantly more communication overhead across different CPU trays.

\begin{table}[]
\centering
\caption{Average processing time per tuple ($T$) under varying NUMA distance. The unit is nanoseconds/tuple.} 
\label{tbl:sub_accuracy}
\resizebox{0.48\textwidth}{!}{%
\begin{tabular}{|l|l|l|l|l|l|}
\hline
\multicolumn{3}{|c|}{Splitter}      & \multicolumn{3}{c|}{Counter}        \\ \hline
From-to      & Measured & Estimated & From-to      & Measured & Estimated \\ \hline
S0-S0(local) & 1612.8   & 1612.8    & S0-S0(local) & 612.3    & 612.3     \\ \hline
S0-S1        & 1666.5  & 1991.1   & S0-S1        & 611.4    & 665.2    \\ \hline
S0-S3        & 1708.2   & 1994.9   & S0-S3        & 623.1   & 665.9    \\ \hline
S0-S4        & 2050.6  & 2923.7   & S0-S4        & 889.9   & 837.9    \\ \hline
S0-S7        & 2371.3  & 3196.4   & S0-S7        & 870.2   & 888.4    \\ \hline
\end{tabular}%
}
\end{table}


To validate the overall effectiveness of our performance model, we show the relative error associated with estimating the application throughput by our analytical model. 
The relative error is defined as $relative\_error =\frac{|R_{meas} - R_{est}|}{R_{meas}}$,
where $R_{meas}$ is the measured application throughput and $R_{est}$ is the estimated application throughput by our performance model for the same application. 

The model accuracy evaluation of all applications under the optimal execution plan on eight CPU sockets is shown in Table~\ref{tbl:accuracy}. 
Overall, our estimation approximates the measurement well for the throughput of all four applications. It is able to produce the optimal execution plan and predict the relative performance quite accurately.

\begin{table}[]
\centering
\caption{Model accuracy evaluation of all applications. The performance unit is K events/sec} 
\label{tbl:accuracy}
\resizebox{0.35\textwidth}{!}{%
\begin{tabular}{|l|l|l|l|l|}
\hline
               & WC       & FD     & SD      & LR     \\ \hline
Measured       & 96390.8  & 7172.5 & 12767.6 & 8738.3 \\ \hline
Estimated      & 104843.3 & 8193.9 & 12530.2 & 9298.7 \\ \hline
Relative error & 0.08     & 0.14   & 0.02    & 0.06   \\ \hline
\end{tabular}%
}
\end{table}

\begin{figure*}[h]
    \centering
	\begin{minipage}[b]{1.01\textwidth}      
	 	\begin{minipage}[c]{0.24\textwidth}  
	    \includegraphics*[width=\textwidth]{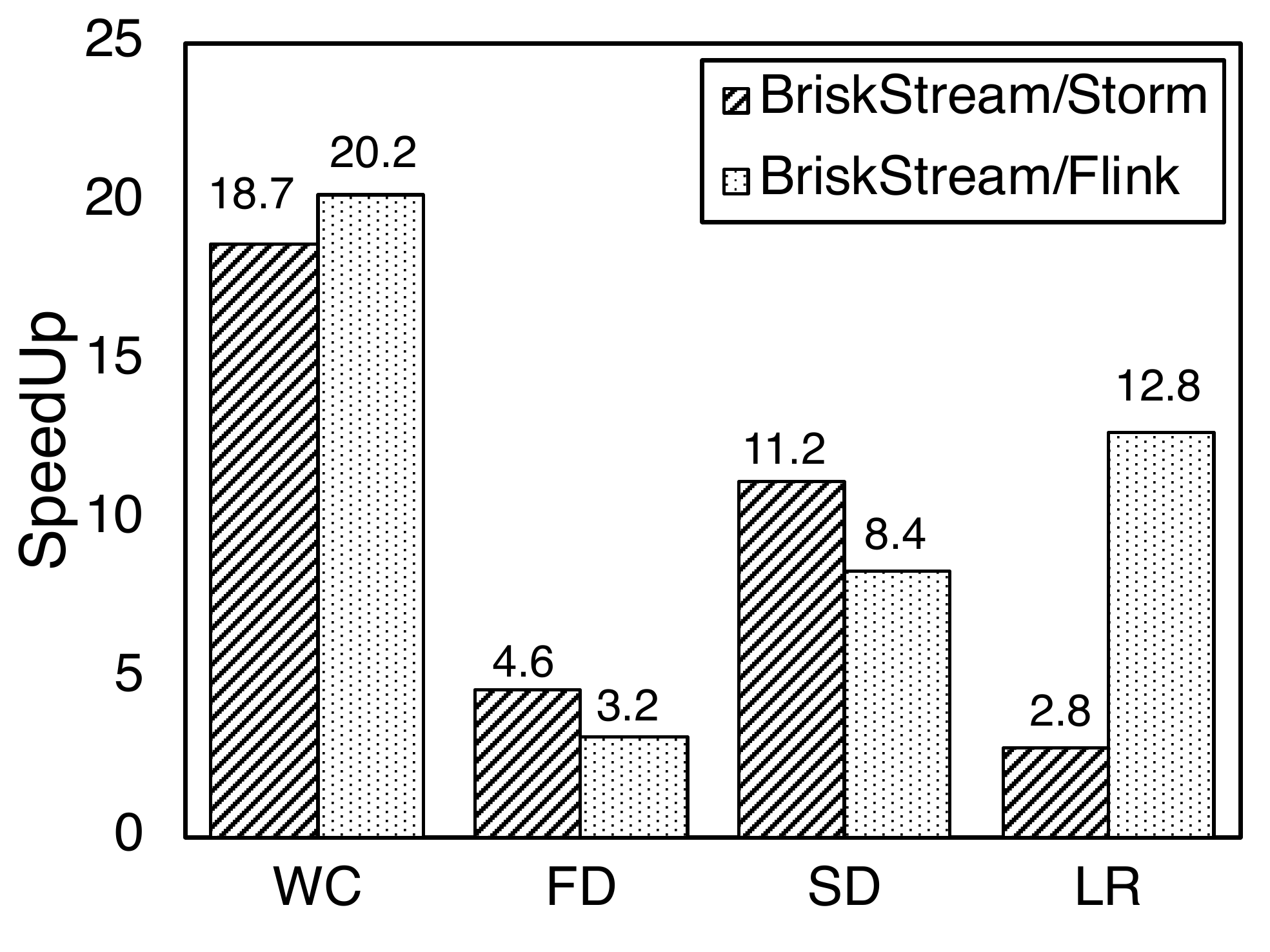}
	    \caption{Throughput speedup.} 
	    \label{fig:end-to-end_throughput} 
	    \end{minipage}
	    \hfill
	    \begin{minipage}[c]{0.25\textwidth} 
	    	\includegraphics*[width=\textwidth]{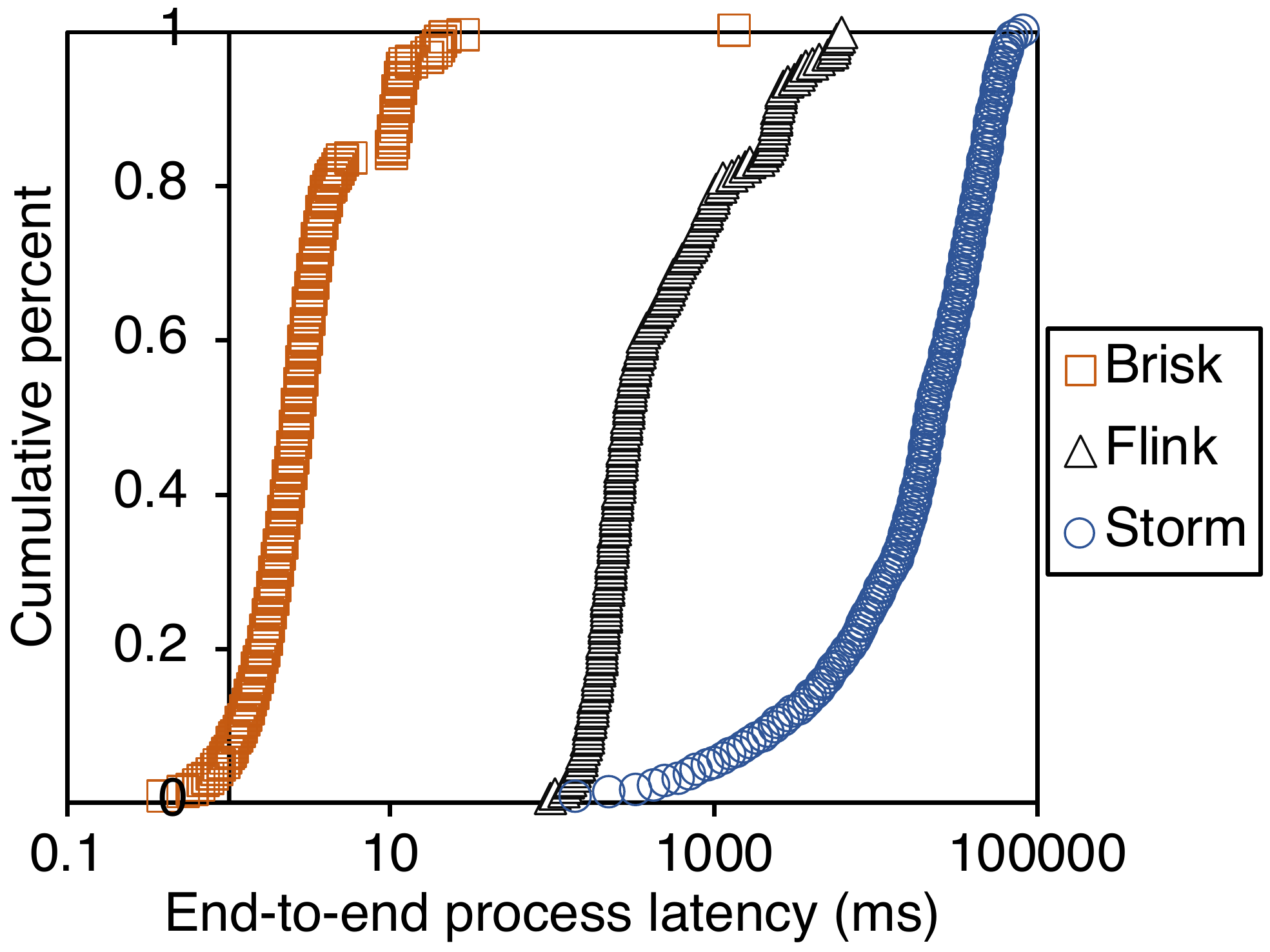} 
	    	\caption{End-to-end latency of WC on different DSPSs.}          
	    	\label{fig:end-to-end_latency} 
	    \end{minipage} 
	    \hfill
	    \begin{minipage}[c]{0.21\textwidth}  
	    \centering
	    \captionof{table}{99-percentile end-to-end latency (ms)}
	    \label{tbl:latency}
	    \resizebox{\textwidth}{!}{%
	    \begin{tabular}{|l|l|l|l|}
	    \hline
	       & \begin{tabular}[c]{@{}l@{}}Brisk\\ Stream\end{tabular} & Storm   & Flink  \\ \hline
	    WC & 21.9                                                   & 37881.3 & 5689.2 \\ \hline
	    FD & 12.5                                                   & 14949.8 & 261.3  \\ \hline
	    SD & 13.5                                                   & 12733.8 & 350.5  \\ \hline
	    LR & 204.8                                                  & 16747.8 & 4886.2 \\ \hline
	    \end{tabular}%
	    }
	    \end{minipage}
	    \hfill
 		\begin{minipage}[c]{0.28\textwidth}  
 	     \includegraphics*[width=\textwidth]{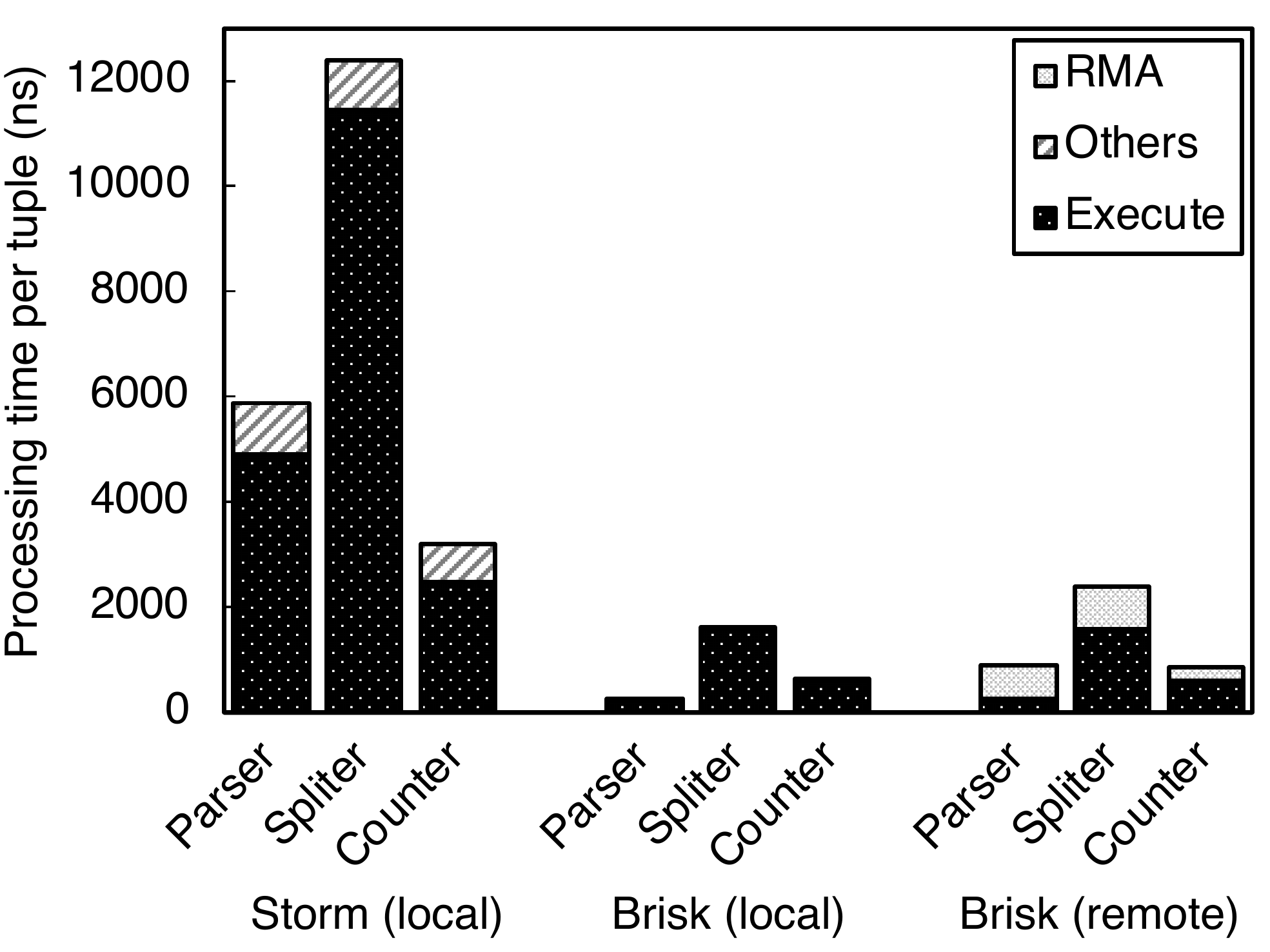}    
 	     \caption{Execution time breakdown.} 
 	     \label{fig:breakdown}     	
 	    \end{minipage}        
	\end{minipage}
	\begin{minipage}[c]{1.01\textwidth}    	    
	 	\begin{minipage}[c]{0.5\textwidth} 
	    \subfloat[Systems]{ 
		    \includegraphics*[width=0.49\textwidth]{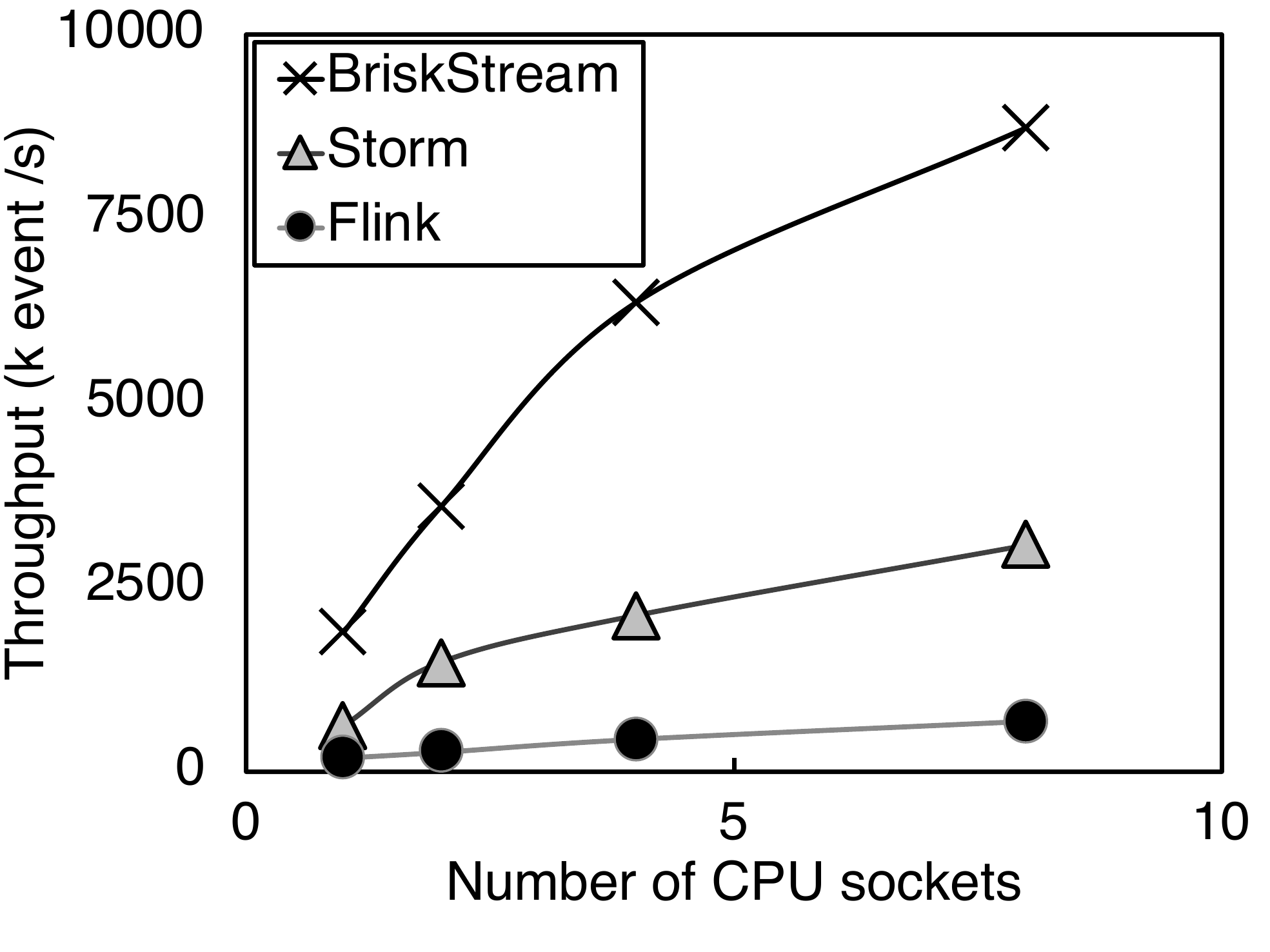}  
		    \label{fig:scalability_lr}     	
	    }
	    \subfloat[Applications]{ 
		    \includegraphics*[width=0.49\textwidth]{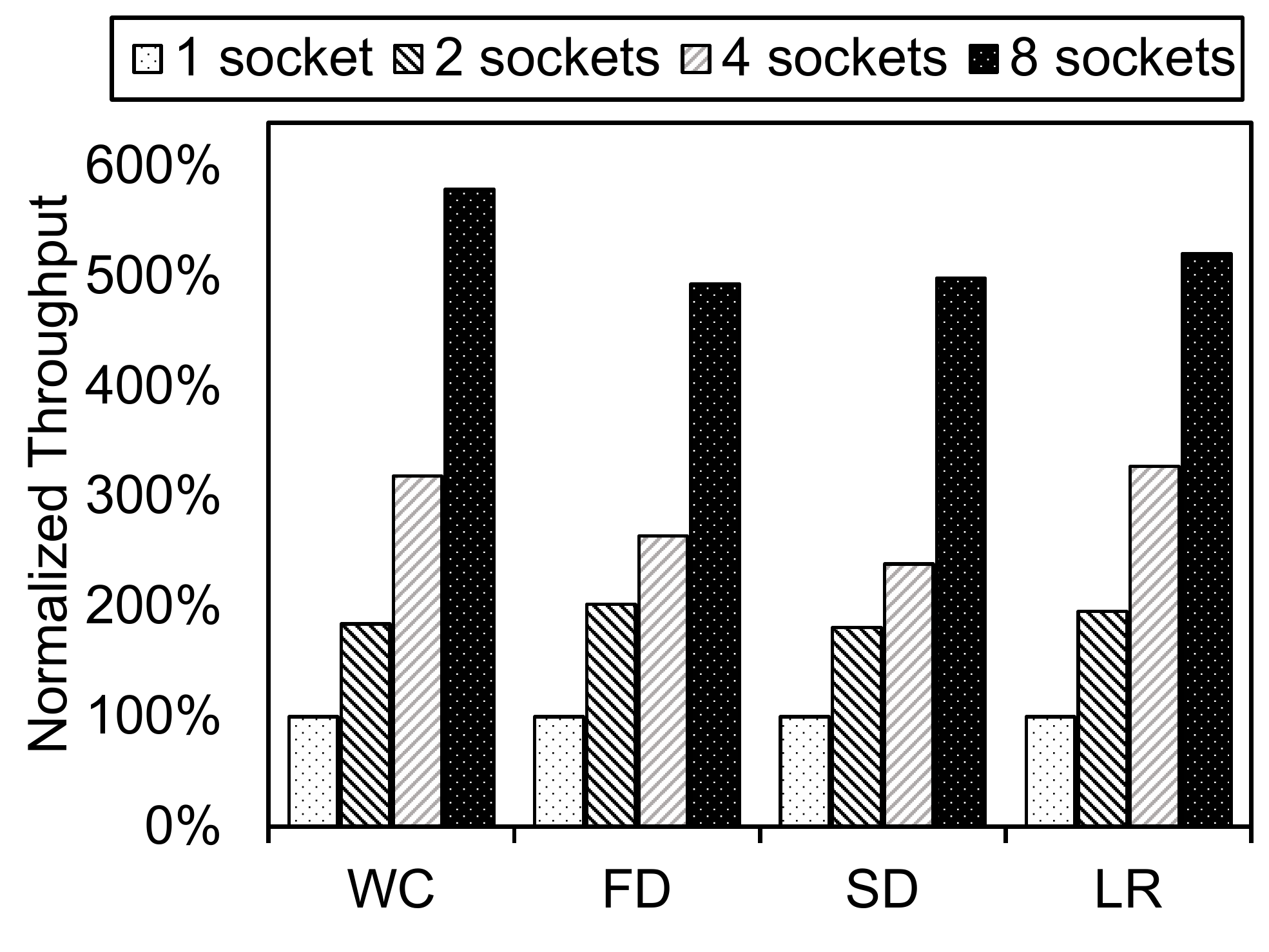}  	   
		    \label{fig:scalability_nus}   
	    } 
	    \caption{Scalability evaluation.}     
	    \end{minipage}   	
	    \hfill
	    \begin{minipage}[c]{0.23\textwidth} 
	    \includegraphics*[width=\textwidth]{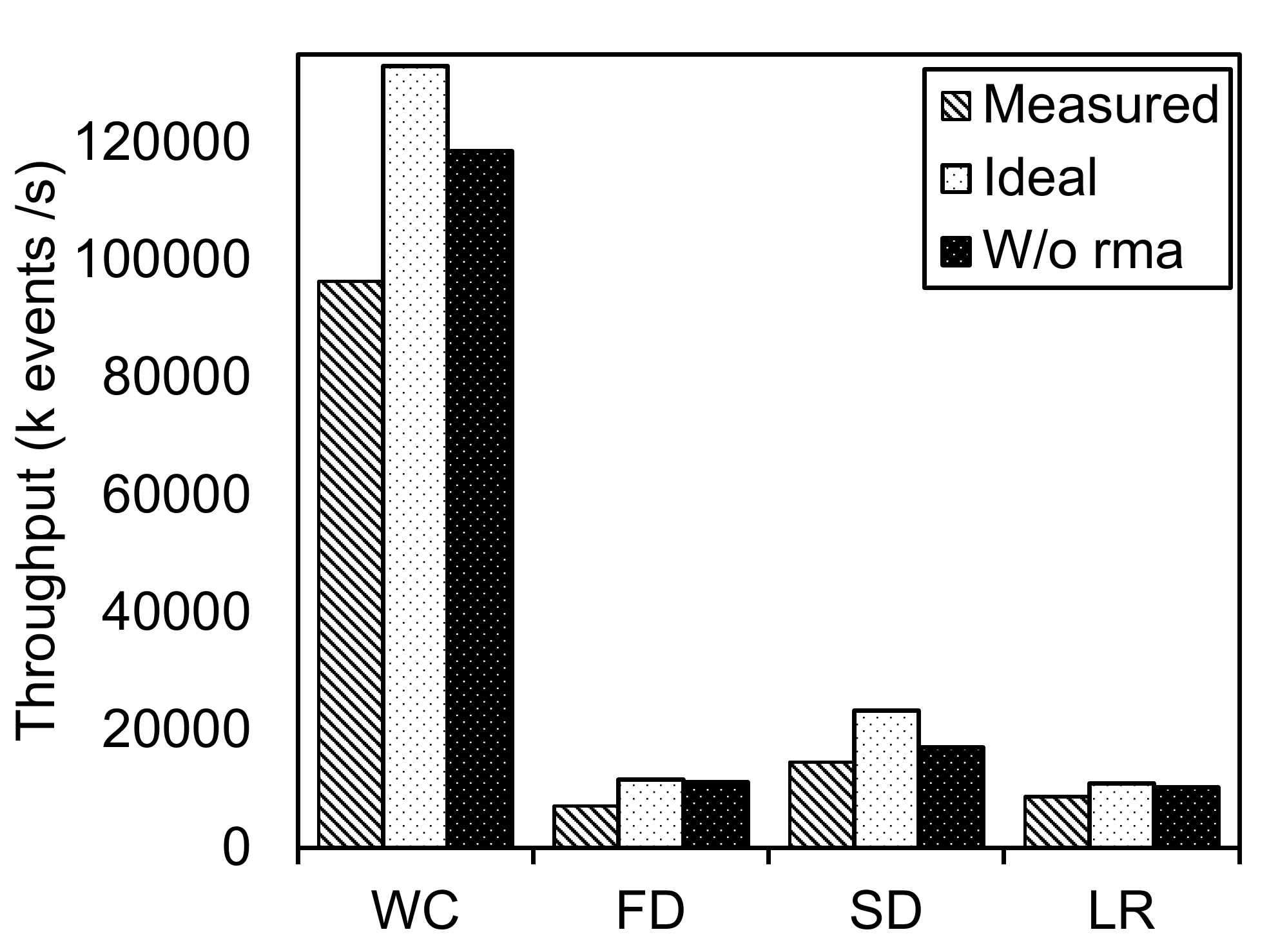}
	    \caption{Gaps to ideal.}   
	    \label{fig:scalability_bound}
	    \end{minipage}	 
 	    \hfill
 		\begin{minipage}[c]{0.26\textwidth}  
 	     \includegraphics*[width=\textwidth]{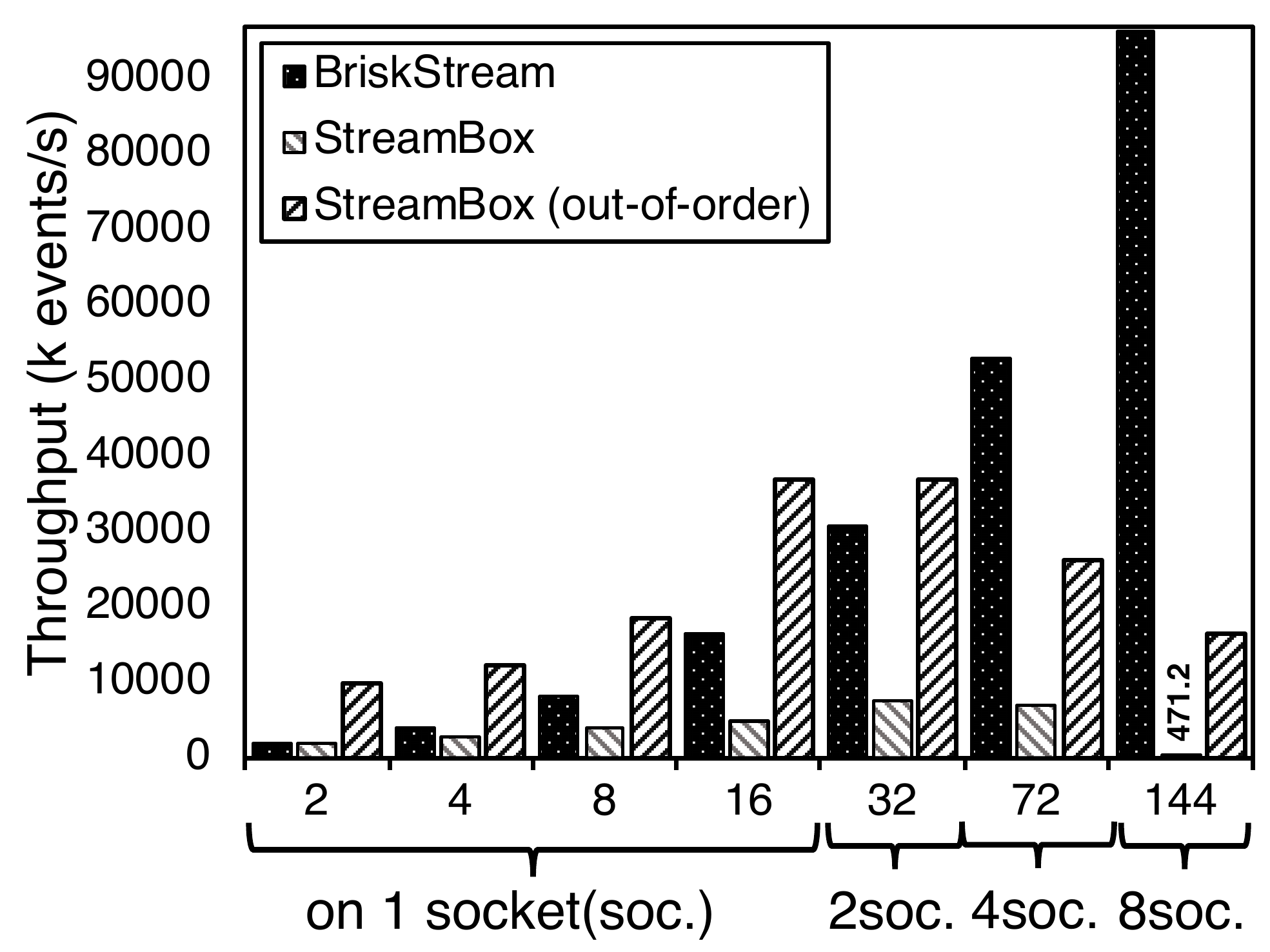}      
 	    \caption{Comparing with StreamBox.} 	
 	    \label{fig:streambox}     
 	    \end{minipage}          
 	\end{minipage}	
\end{figure*}

\subsection{Evaluation of Execution Efficiency}
\label{subsec:design}
This section shows that BriskStream significantly outperforms existing DSPSs on shared-memory multicores.
We compare \system with two open-sourced DSPSs including Apache Storm (version 1.1.1) and Flink (version 1.3.2). 
For a better performance, we disable the fault-tolerance mechanism in all comparing systems. 
We use Flink with NUMA-aware configuration (i.e., one task manager per CPU socket), and as a sanity check, we have also tested Flink with a single task manager, which shows even worse performance. 
We also compare \system with StreamBox, a recent single-node DSPS on share-memory multi-core architectures at the end of this section.

\textbf{Throughput and latency comparison.}
Figure~\ref{fig:end-to-end_throughput} shows the significant throughput speedup of BriskStream compared to Storm and Flink. 
Overall, Storm and Flink show comparable throughput for three applications including WC, FD and SD. 
Flink performs poorly for LR compared to Storm. A potential reason is that Flink requires additional stream merger operators (implemented as the co-flat map) that merges multiple input streams before feeding to an operator with multi-input streams (commonly found in LR). Neither Storm nor BriskStream has such additional overhead.

Following the previous work~\cite{adaptivebatch}, we define the end-to-end latency of a streaming workload as the duration between the time when an input event enters the system and the time when the results corresponding to that event is generated. 
We compare the end-to-end process latency among different DSPSs on Server A.
Figure~\ref{fig:end-to-end_latency} shows the detailed CDF of end-to-end processing latency of WC comparing different DSPSs and Table~\ref{tbl:latency} shows the overall 99-percentile end-to-end processing latency comparison of different applications.
The end-to-end latency of \system is significantly smaller than both Flink and Storm.

\textbf{Per-tuple execution time breakdown.}
To better understand the source of performance improvement, we show the per-tuple execution time breakdown by comparing \system and Storm. 
Figure~\ref{fig:breakdown} shows the breakdown of all non-source operators of WC, which we use as the example application in this study.
We perform analysis in two groups:
\emph{local} stands for allocating all operators to the same socket, and 
\emph{remote} stands for allocating each operator max-hop away from its producer to examine the cost of RMA.

In the local group, we compare execution efficiency between BriskStream and Storm. 
The ``others'' overhead of each operator is commonly reduced to about $10\%$ of that of Storm. 
The function execution time is also significantly reduced to only $5\sim24\%$ of that of Storm.
There are two main reasons for this improvement. 
First, the instruction cache locality is significantly improved due to much smaller code footprint. 
In particular, our further profiling results reveal that BriskStream is no longer front-end stalls dominated (less than $10\%$), while Storm and Flink are (more than $40\%$).
Second, our ``jumbo tuple'' design eliminates duplicate metadata creation and effectively amortizes the communication queue access overhead.

In the remote group, we compare the execution of the same operator in BriskStream with or without remote memory access overhead. 
In comparison with the locally allocated case, the total round trip time of an operator is up to 9.4 times higher when it is remotely allocated to its producer. In particular, Parser has little in computation but has to pay a lot for remote memory access overhead ($T^{e} << T^{f}$). 
The significant varying processing capability of the same operator when it is under different placement plan reaffirms the necessity of our RLAS optimization.

Another takeaway is that \emph{Execute} in Storm is much larger than \emph{RMA}, which means $T^{e} >> T^{f}$ and NUMA effect may have a minor impact in its plan optimization. 
In contrast, \system significantly reduces $T^{e}$ (discussed in Section~\ref{sec:impl}) and the NUMA effect\tony{, as a result of improving efficiency of other components,} becomes a critical issue to optimize. 
In the future, on one hand, $T^{e}$ may be further reduced with more optimization techniques deployed. On the other hand, servers may scale to even more CPU sockets (with potentially larger max-hop remote memory access penalty). 
We expect that those two trends make the NUMA effect continues to play an important role in optimizing streaming computation on shared-memory multicores.


\textbf{Evaluation of scalability on varying CPU sockets.}
Our next experiment shows that BriskStream scales effectively as we increase the numbers of sockets. 
RLAS is able to optimize the execution plan under a different number of sockets enabled. 
Figure~\ref{fig:scalability_lr} shows the better scalability of BriskStream than existing DSPSs on multi-socket servers by taking LR as an example. 
Unmanaged thread interference and unnecessary remote memory access penalty prevent existing DSPSs from scaling well on the modern multi-sockets machine.
We show the scalability evaluation of different applications of BriskStream in Figure~\ref{fig:scalability_nus}. 
There is an almost linear scale up from 1 to 4 sockets for all applications. 
However, the scalability becomes poor when more than 4 sockets are used. 
This is because of a significant increase of RMA penalty between upper and lower CPU tray. In particular, RMA latency is about two times higher between sockets from different tray than the other case.

To better understand the effect of RMA overhead during scaling, we compare the theoretical bounded performance without RMA (denoted as ``W/o rma'') and ideal performance if the application is linearly scaled up to eight sockets (denoted as ``Ideal'') in Figure~\ref{fig:scalability_bound}. 
The bounded performance is obtained by evaluating the same execution plan on eight CPU sockets by substituting RMA cost to be zero.
There are two major insights from Figure~\ref{fig:scalability_bound}.
First, theoretically removing RMA cost (i.e., ``W/o rma'') achieves $89\sim95\%$ of the ideal performance, and it hence confirms that the significant increase of RMA cost is the main reason that \system is not able to scale linearly on 8 sockets.
Second, we still need to improve the parallelism and scalability of the execution plan to achieve optimal performance even without RMA.

\textbf{Comparing with single-node DSPS.}
Streambox~\cite{StreamBox} is a recently proposed DSPS based on a morsel-driven like execution model -- a different processing model to \system.
We compare BriskStream with StreamBox using WC as an example. 
Results in Figure~\ref{fig:streambox} demonstrate that BriskStream outperforms StreamBox significantly regardless of the number of CPU cores used in the system. 
Note that, StreamBox focuses on solving out-of-order processing problem, which requires more expensive processing mechanisms such as locks and container design. 
Due to a different system design objective, \system currently does not provide ordered processing guarantee and consequently does not bear such overhead.

For a better comparison, we modify StreamBox to disable its order-guaranteeing feature, denoted as StreamBox (out-of-order), so that tuples are processed out-of-order in both systems. 
Despite its efficiency at smaller core counts, it scales poorly when multiple sockets are used. 
There are two main reasons. 
\tony{
First, StreamBox relies on a centralized task scheduling/distribution mechanism with locking primitives, which brings significant overhead for more CPU cores. 
This could be a limitation inherited from adopting morsel-driven execution model in DSPSs -- essentially it trades off the reduced pipeline parallelism for lower operator communication overhead, which we defer as a future work to investigate in more detail.
}
Second, WC needs the same word being counted by the same counter, which requires a data shuffling operation in StreamBox. Such data shuffling operation introduces significant remote memory access to StreamBox.
We compare their NUMA overhead during execution using Intel Vtune Amplifier~\cite{Vtune}.
We observe that, under 8 sockets (144 cores), BriskStream issues in average 0.09 cache misses served remotely per k events (misses/k events), which StreamBox's has 6 misses/k events. 


 \begin{figure*}[h]
     \centering  
 	\begin{minipage}{1.01\textwidth}  
 		 \hfill
 		\begin{minipage}[c]{0.28\textwidth}
 		    \centering
 		 	\includegraphics*[width=\textwidth]{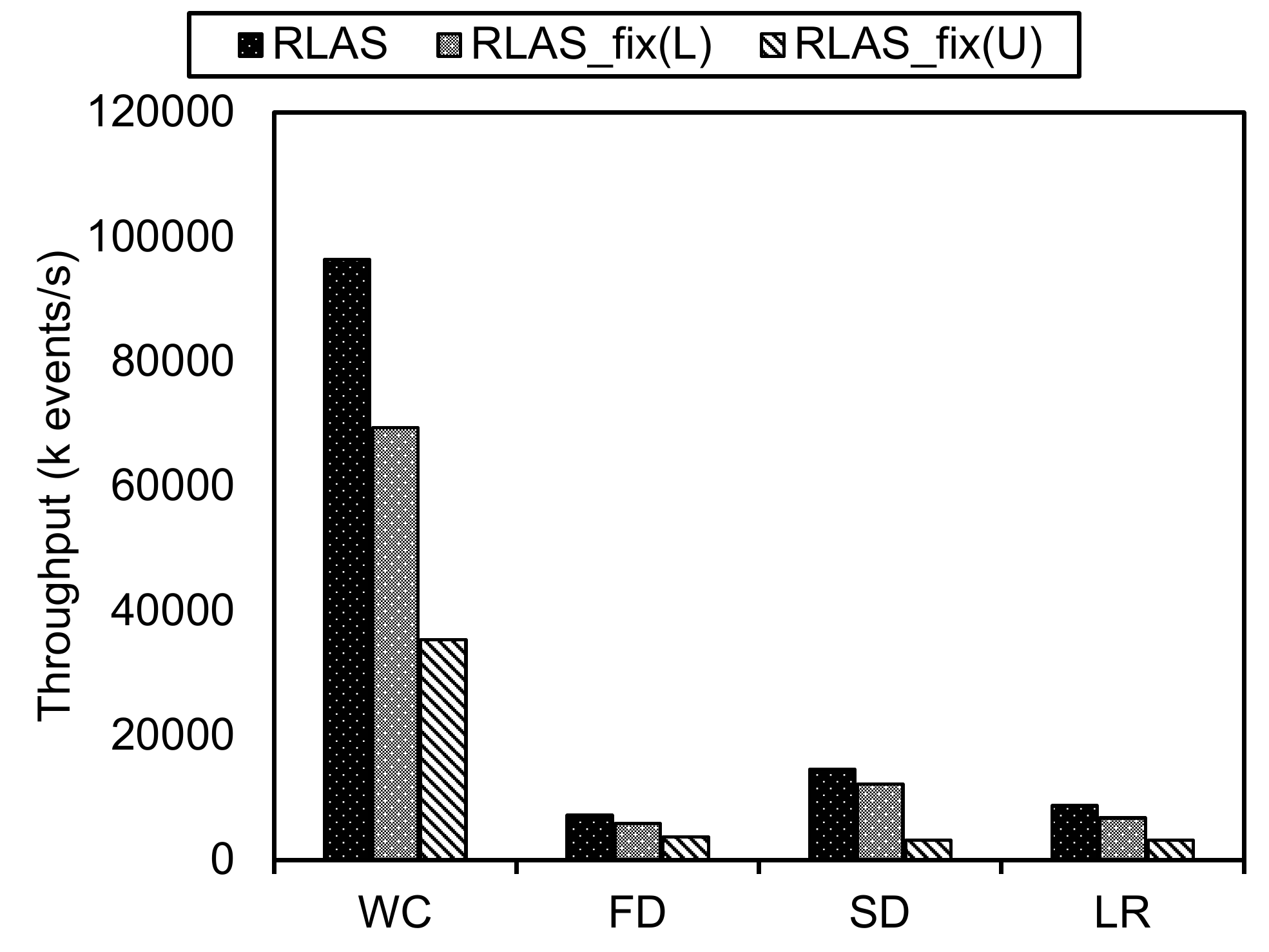}
 		    \caption{RLAS w/ and w/o considering varying RMA cost.}
 		    \label{fig:RLAS}
 		 \end{minipage}	  		 
 		 \begin{minipage}[c]{.26\textwidth}
 		     \centering
 		     \captionof{table}{Placement strategies}
 		     \label{fig:technique}
 		     \includegraphics[width=\textwidth]{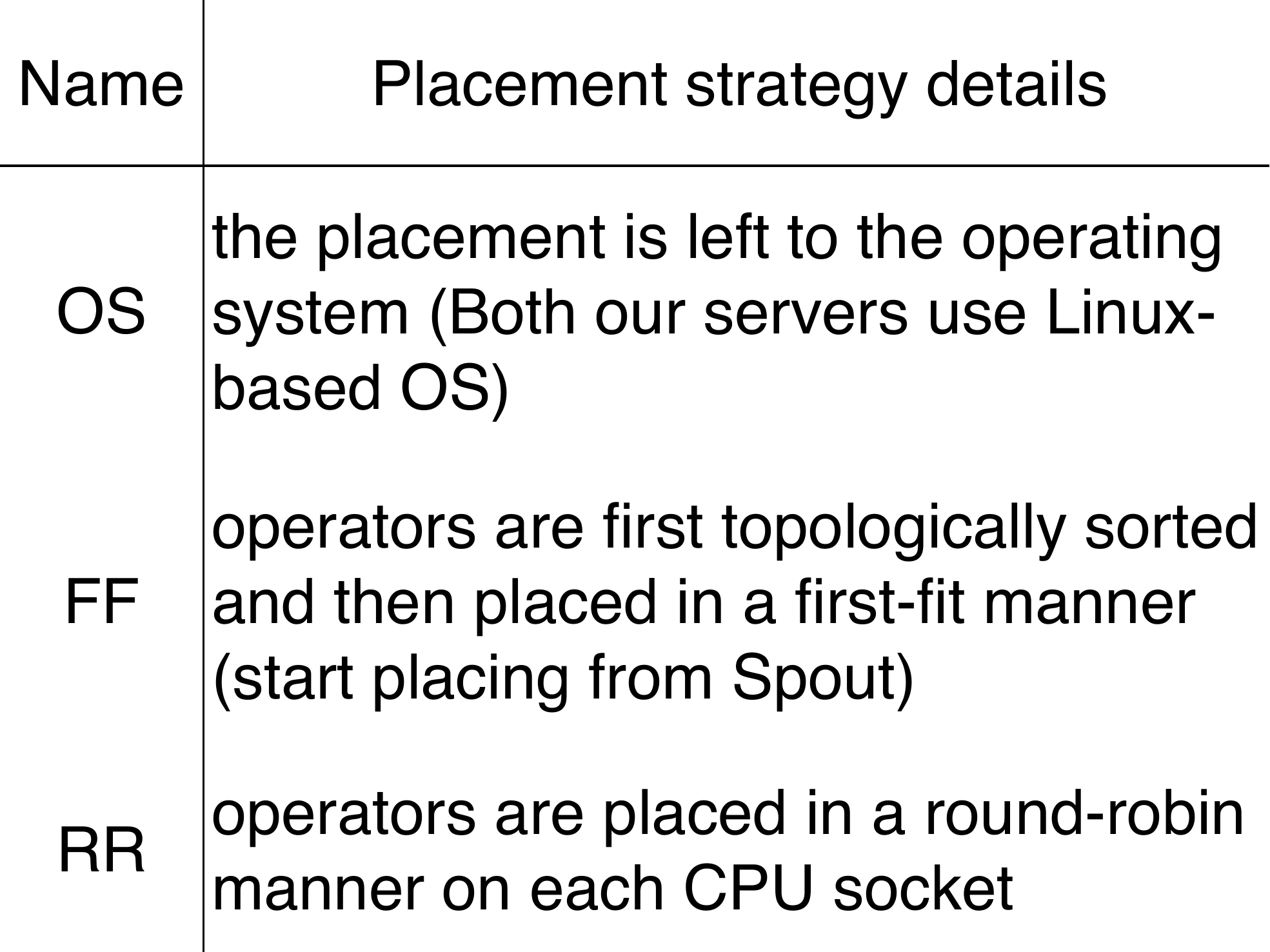} 
 		 \end{minipage}
 		 \begin{minipage}[c]{.44\textwidth}
 		  \subfloat[Server A]{%
 		  \includegraphics[width=0.5\textwidth]{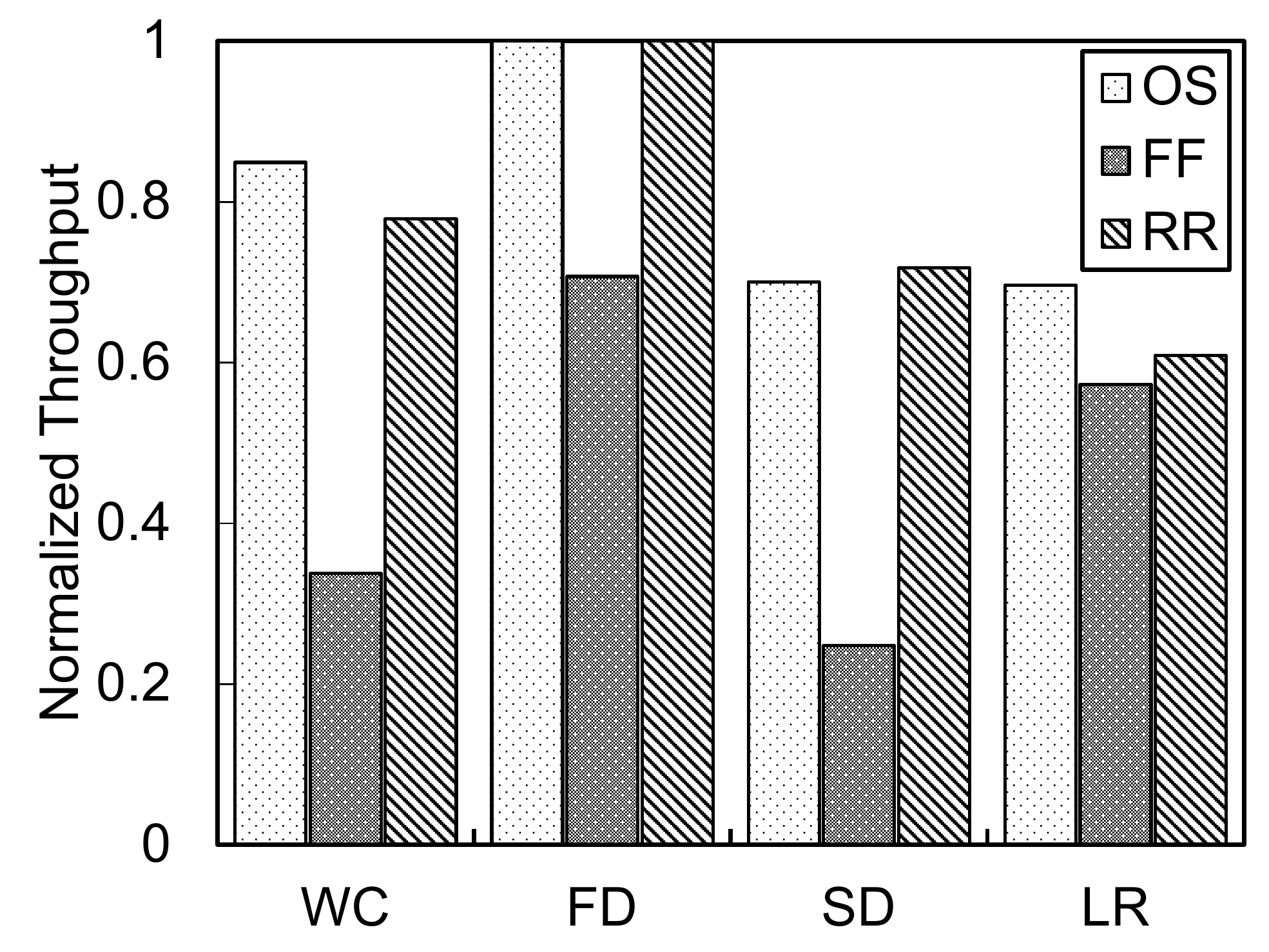}   \label{fig:Compare1}
 		 }      
 		 \subfloat[Server B]{%
 		 \includegraphics[width=0.5\textwidth]{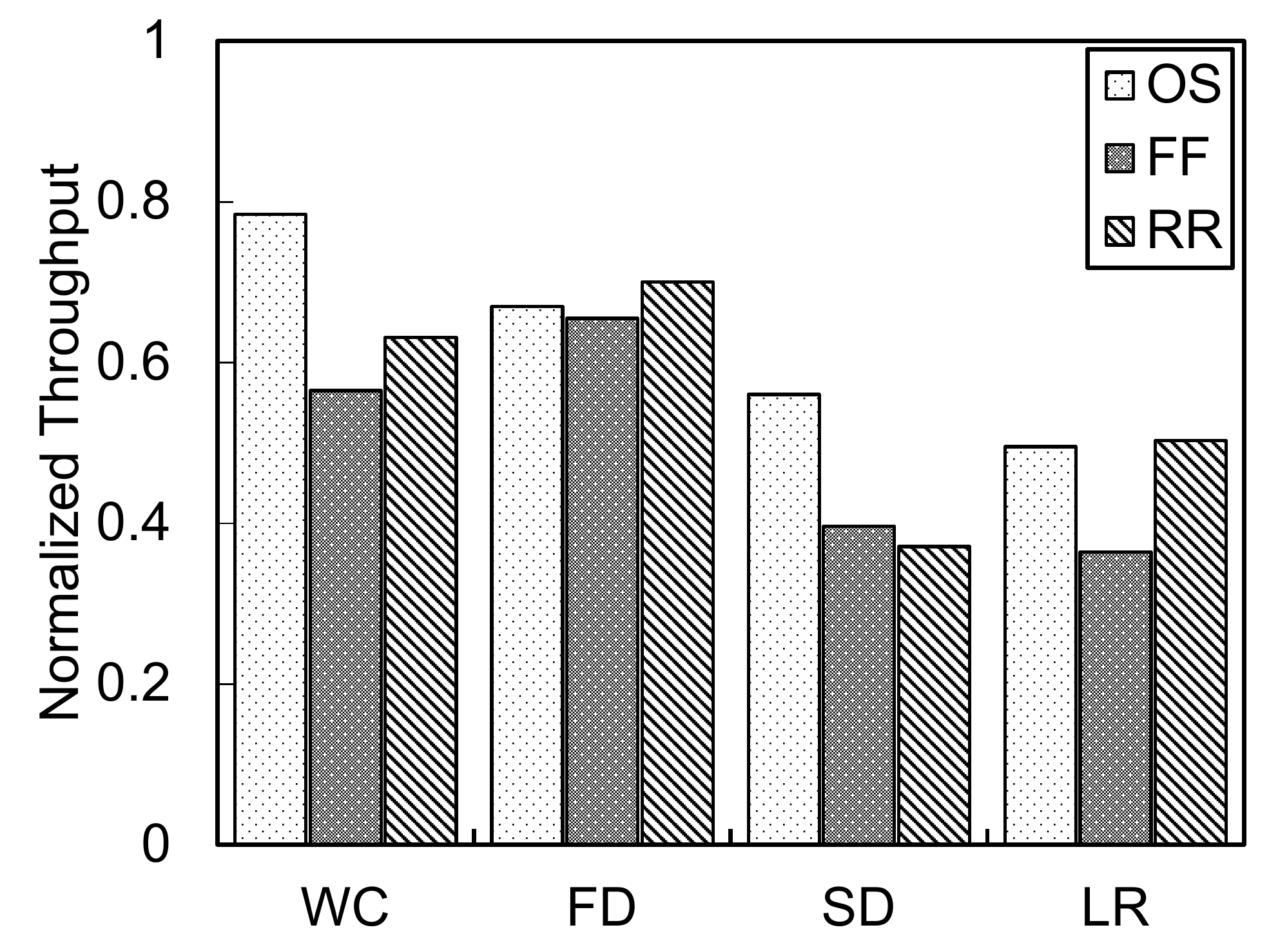}    \label{fig:Compare2}
 		 }
 		\caption{Placement strategy comparison under the same replication configuration.}                   
 		\label{fig:Compare} 
 		\end{minipage}    
 	\end{minipage}  
 \end{figure*}
   
\subsection{Evaluation of RLAS algorithms}
\label{subsec:rlas}
In this section, we study the effectiveness of RLAS optimization and compare it with competing techniques.




\textbf{The necessity of considering varying processing capability.}
To gain a better understanding of the importance of relative-location awareness, 
we consider an alternative algorithm that utilizes the same searching process of RLAS but assumes each operator has a fixed processing capability. 
Such an approach essentially falls back to the original RBO model~\cite{Viglas:2002:RQO:564691.564697}, and is also similar to some previous works~\cite{Deployment,Khandekar:2009:COS:1656980.1657002}.
In our context, we need to fix $T^f$ of each operator to a constant value. 
We consider two extreme cases.
First, the lower bound case, namely $RLAS\_fix(L)$, assumes each operator pessimistically always includes remote access overhead. That is, $T^f$ is calculated by anti-collocating an operator to all of its producers. 
Second, the upper bound case, namely $RLAS\_fix(U)$, completely ignores RMA, and $T^f$ is set to 0 regardless the relative location of an operator to its producers.

The comparison results are shown in Figure~\ref{fig:RLAS}. 
RLAS shows a 19\% $\sim$ 39\% improvement over $RLAS\_fix(L)$. 
We observe that $RLAS\_fix(L)$ often results in smaller replication configuration of the same application compared to RLAS and hence underutilizes the underlying resources. This is because it \emph{over-estimates} the resource demand of operators that are collocated with producers. 
Conversely, $RLAS\_fix(U)$ \emph{under-estimates} the resource demands of operators that are anti-collocated and misleads the optimization process to involve severely thread interference. 
Over the four workloads, RLAS shows a 119\% $\sim$ 455\% improvement over $RLAS\_fix(U)$.
 

\textbf{Comparing different placement strategies.}
We now study the effect of different placements under the same replication configuration. 
In this experiment, the replication configuration is fixed to be the same as the optimized plan generated by RLAS and only the placement is varied under different techniques. 
Three alternative placement strategies are shown in Table~\ref{fig:technique}.
Both FF~\cite{T-storm} and RR~\cite{R-storm} are enforced to guarantee resource constraints as much as possible.
In case these strategies cannot find any plan satisfying resource constraints, they will gradually relax constraints until a plan is obtained.
We also configure external input rate ($I$) to just overfeed the system on Server A, 
and using the same $I$ to test on Server B. This allows us to examine the system capacity of different servers.
The results are shown in Figure~\ref{fig:Compare}.
There are two major takeaways.

First, RLAS generally outperforms other placement techniques on both two servers. 
FF can be viewed as a minimizing traffic heuristic-based approach as it greedily allocates neighbor operators (i.e., directly connected) together due to its topologically sorting step. 
Several related studies~\cite{T-storm, Adaptive} adopt a similar approach of FF in dealing with operator placement problem in the distributed environment. 
However, it performs poorly, because we find that during its searching for optimal placements, it often falls into ``not-able-to-progress" situation as it cannot allocate the current operator into any of the sockets because of the violation of resource constraints. This is due to its greedy nature that leads to a local optimal state.
Then, it has to relax the resource constraints and repack the whole topology, which often ends up with oversubscribing of a few CPU sockets.
The major drawback of RR is that it does not take remote memory communication overhead into consideration, and the resulting plans often involve unnecessary cross-socket communication. 

Second, RLAS performs generally better on Server B.
We observe that Server B is underutilized for all applications under the given testing input loads. This indicates that although the total computing power (aggregated CPU frequency) of Server A is higher, its maximum attainable system capacity is actually smaller.
As a result, RLAS chooses to use only a subset of the underlying hardware resource of Server B to achieve the maximum application throughput. In contrast, other heuristic based placement strategies unnecessarily involve more RMA cost by launching operators to all CPU sockets.

\begin{figure*}    
   	\begin{minipage}{1.01\textwidth}
    	 \begin{minipage}[c]{0.33\textwidth}
    	     \centering
    	  	\includegraphics*[width=\textwidth]{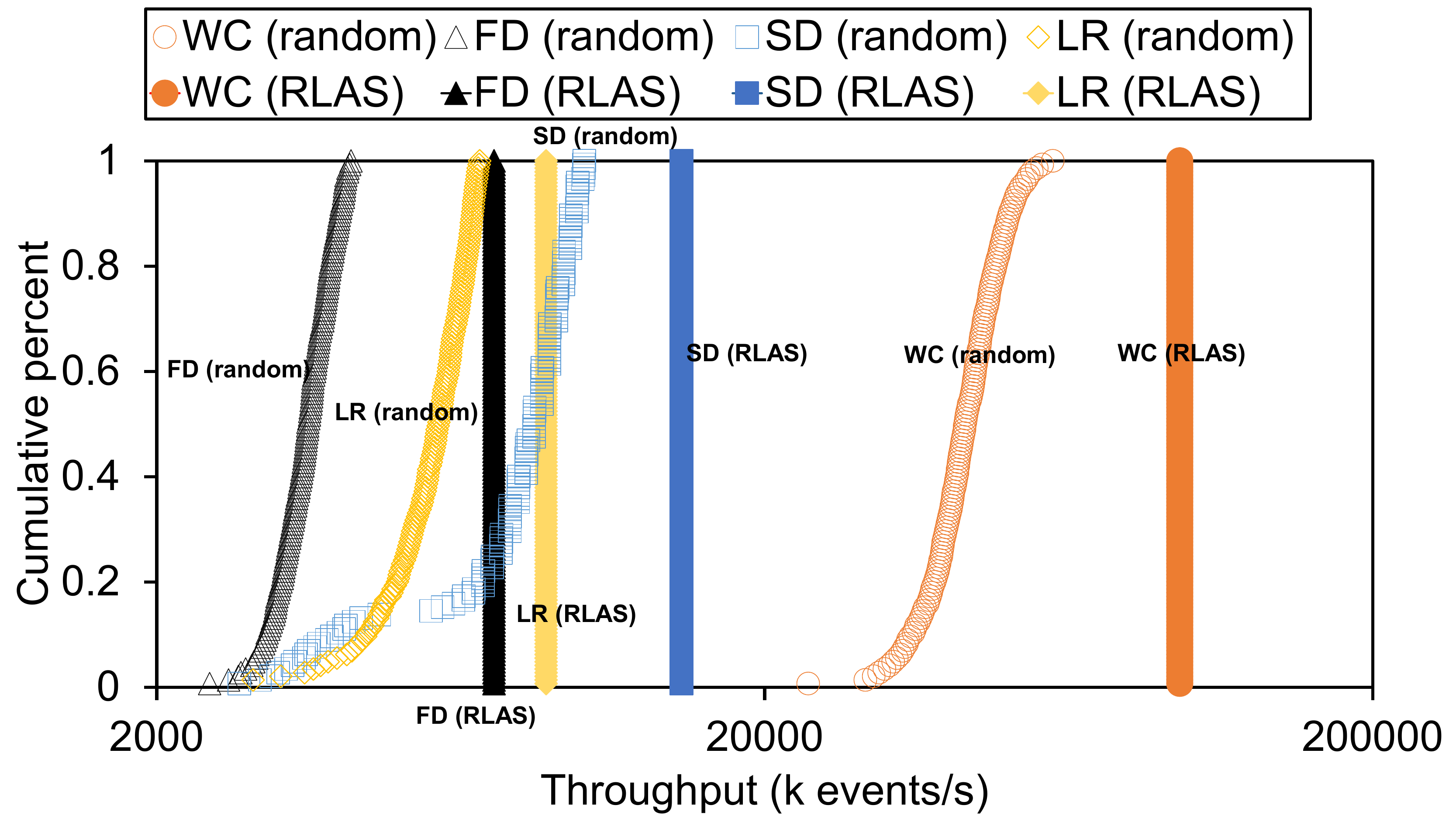}  
    	     	\caption{CDF of random plans.}
    	      \label{fig:modeleva1}    
      	\end{minipage}    
      	\hfill	
   		\begin{minipage}[c]{.44\textwidth}
   	    \includegraphics*[width=\textwidth]{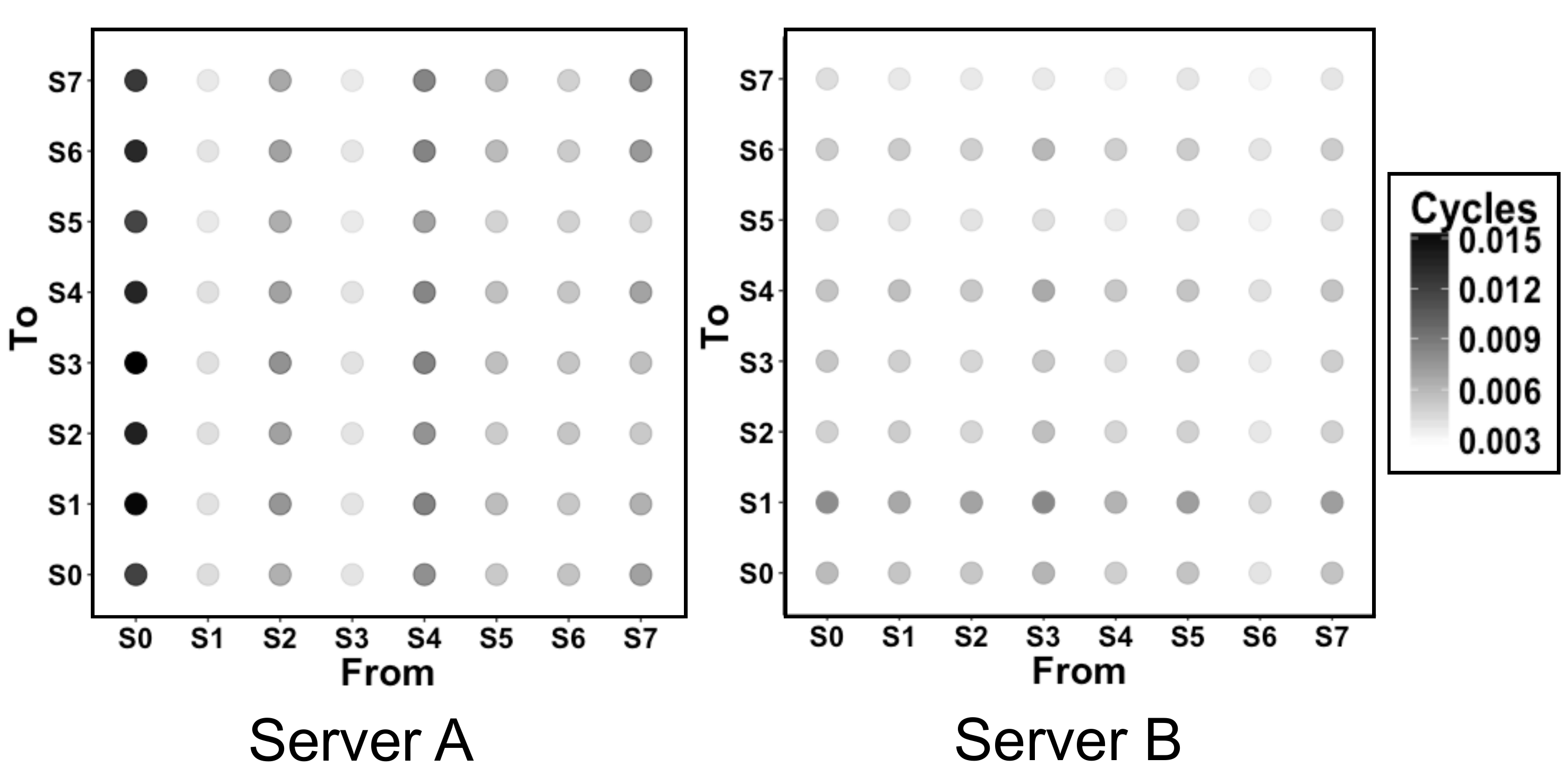}  
   	    \caption{Communication pattern matrices of WC on two Servers.}
   	    \label{fig:commPattern}   	
   	   \end{minipage}
   	   \hfill
 	   \begin{minipage}[c]{0.22\textwidth}  
		\captionof{table}{Runtime of the optimization process}
		\label{tbl:r}
		\resizebox{0.9\textwidth}{!}{%
		\begin{tabular}{|l|l|l|}
		\hline
		r  & throughput & \begin{tabular}[c]{@{}l@{}}runtime\\ (sec)\end{tabular} \\ \hline
		1  & 10140.2    & 93.4                                                    \\ \hline
		3  & 10079.5    & 48.3                                                    \\ \hline
		5  & 96390.8    & 23.0                                                      \\ \hline
		10 & 84955.9    & 46.5                                                    \\ \hline
		15 & 77773.6    & 45.3                                                    \\ \hline
		\end{tabular}%
		}
 	\end{minipage}   	    
   	 \end{minipage}	
\end{figure*}

\textbf{Correctness of heuristics.}
Due to a very large search space, it is almost impossible to examine all execution plans of our test workloads to verify the effectiveness of our heuristics. 
Instead, we utilize Monte-Carlo simulations by generating 1000 random execution plans, and compare against our optimized execution plan. 
Specifically, the replication level of each operator is randomly increased until the total replication level hits the scaling limit. All operators (incl. replicas) are then randomly placed. 
Results of Figure~\ref{fig:modeleva1} show that none of the random plans is better than RLAS. It demonstrates that random plans hurt the performance in a high probability due to the huge optimization space.

\tonyy{
We further observe two properties of optimized plans of RLAS, which are also found in randomly generated plans with relatively good performance.
First, operators of FD and LR are completely avoided being remotely allocated across different CPU-tray to their producers. 
This indicates that the RMA overhead, especially from the costly communications across CPU trays, should be aggressively avoided in these two applications. 
Second, resources are well utilized for high throughput optimizations in RLAS. Most operators (incl. replicas) end up with being ``just fulfilled'', i.e., $\overline{r_o}= r_o = r_i$. This effectively reveals the shortcoming of existing heuristics based approach -- maximizing an operator's performance may be worthless or even harmful to the overall system performance as it may already overfeed its downstream operators. Further increasing its performance (e.g., scaling it up or making it allocated together with its producers) is just a waste of the precious computing resource.
}

\textbf{Communication pattern.}
In order to understand the impact of different NUMA architectures on RLAS optimization, we show communication pattern matrices of running WC with an optimal execution plan in Figure~\ref{fig:commPattern}. The same conclusion applies to other applications and hence omitted.
Each point in the figure indicates the summation of data fetch cost (i.e., $T^{f}$) of all operators from the x-coordinate ($S_i$) to y-coordinate ($S_j$).
The major observation is that the communication requests are mostly sending from one socket (S0) to other sockets in Server A, and they are, in contrast, much more uniformly distributed among different sockets in Server B. 
The main reason is that the remote memory access bandwidth is almost identical to local memory access in Server B thanks to its glue-assisted component as discussed in Section~\ref{sec:background}, and operators are hence more uniformly placed at different sockets.


\textbf{Varying the compression ratio ($r$).}
RLAS allows to compress the execution graph (with a ratio of $r$) to tune the trade-off between optimization granularity and searching space.
We use WC as an example to show its impact as shown in Table~\ref{tbl:r}. Similar trend is observed in other three applications.
Note that, a compressed graph contains heavy operators (multiple operators grouped into one), which may fail to be allocated and requires re-optimization. This procedure introduces more complexity to the algorithm, which leads to higher runtime of the optimization process as shown in Table~\ref{tbl:r}.
Due to space limitation, a detailed discussion is presented in Appendix~\ref{subsec:discuss}.

%


\subsection{Factor Analysis}
\label{subsec:factor}
\begin{figure}
\centering
    \includegraphics*[width=0.45\textwidth]{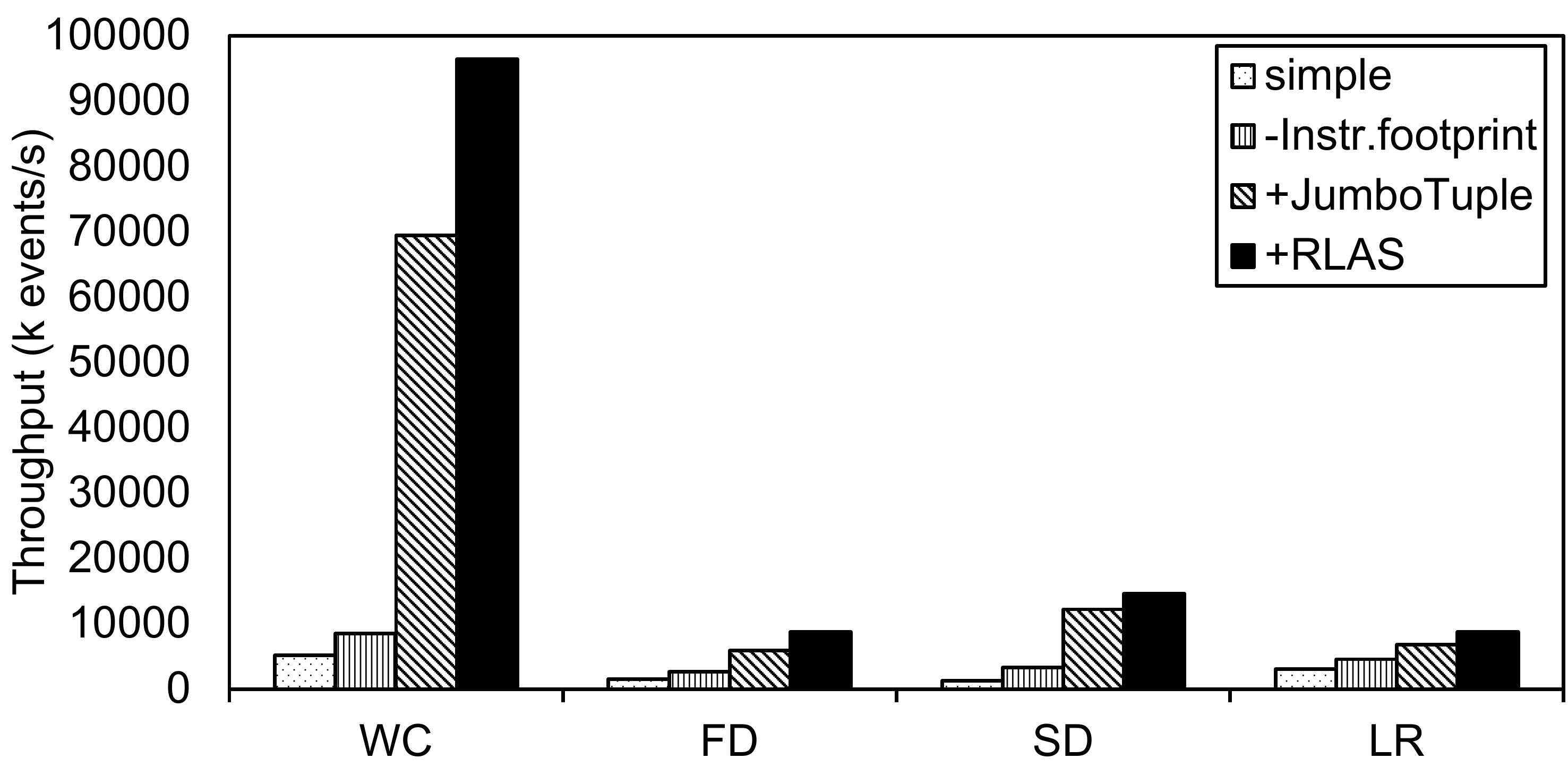}  
    \caption{A factor analysis for BriskStream. Changes are added left to right and are cumulative.}
    \label{fig:factor}
\end{figure}

To understand the details in the overheads and benefits of various aspects of BriskStream, we show a factor analysis in Figure~\ref{fig:factor} that highlights the key factors for performance. 
\emph{Simple} refers to running Storm directly on shared-memory multicores. 
\emph{-Instr.footprint} refers to \system with much smaller instruction footprint and avoiding unnecessary/duplicate objects as described in Section~\ref{subsec:impl1}. 
\emph{+JumboTuple} further allows \system to reduce the cross-operator communication overhead as described in Section~\ref{subsec:impl2}. 
In the first three cases, the system is optimized under $RLAS\_fix (L)$ scheme without considering \emph{varying} RMA cost.
\emph{+RLAS} adds our NUMA aware execution plan optimization as described in Section~\ref{sec:rlas}. 
The major takeaways from Figure~\ref{fig:factor} are that jumbo tuple design is important to optimize existing DSPSs on shared-memory multicore architecture and our RLAS optimization paradigm is critical for DSPSs to scale different applications on modern multicores environment addressing NUMA effect.

\compact
\section{Related Work}
\label{sec:related}
\emph{Database optimizations on scale-up architectures:}
Scale-up architectures have brought many research challenges and opportunities for in-memory data management, as outlined in recent surveys~\cite{inmemorydbsurvey, inmemorybigdata}. 
There have been studies on optimizing the instruction cache performance~\cite{Zhou:2004:BDO:1007568.1007592, Harizopoulos:2006:IIC:1166074.1166079}, the memory and cache performance~\cite{AilamakiVLDB99, 1410200, databaseNewBottleneck, balkesen13}, many-core parallelism in a single chip~\cite{Cheng:2017:SMH:3132847.3132916, Jha:2015:IMM:2735703.2735704} and NUMA~\cite{Leis:2014:MPN:2588555.2610507,LiPMRL13,Deployment,Porobic:2012:OHI:2350229.2350260,psaroudakis2015scaling,Psaroudakis2016}. 
Psaroudakis et al.~\cite{psaroudakis2015scaling,Psaroudakis2016} developed an adaptive data placement algorithm that can track and resolve utilization imbalance across sockets. 
However, it does not solve the problem we address.
In particular, the placement strategy such as RR balances resource utilization among CPU sockets, but shows suboptimal performance in our experiments.
Leis et al.~\cite{Leis:2014:MPN:2588555.2610507} proposed a novel morsel-driven query execution model which integrates both NUMA-awareness and fine grained task-based parallelism. 
A similar execution model is adopted in StreamBox~\cite{StreamBox}, which we compared in our experiments. The results confirm the superiority of \system in addressing NUMA effect. 

\emph{Data stream processing systems (DSPSs):}
DSPSs have attracted a great amount of research effort. 
A number of systems have been developed, for example, TelegraphCQ~\cite{TelegraphCQ}, Borealis~\cite{Borealis}, IBM System S~\cite{system-s} and the more recent ones including Storm~\cite{Storm}, Flink~\cite{flink} and Heron~\cite{heron}. 
However, most of them targeted at the distributed environment, and little attention has been paid to the research on DSPSs on the modern multicore environment.
A recent patch on Flink~\cite{NUMA_flink} tries to make Flink a NUMA-aware DSPS. However, its current heuristic based round-robin allocation strategy is not sufficient to make it scale on large multicores as our experiment shows.
Previous work~\cite{profile} gave a detailed study on the insufficiency of two popular DSPSs (i.e., Storm and Flink) running on modern multi-core processors. It proposed a heuristic-based algorithm to deploy stream processing on NUMA-based machines. However, the heuristic does not take relative-location awareness into account. It may not always be efficient for different workloads. In contrast, BriskStream provides a model-guided approach that automatically determines the optimal operator parallelism and placement addressing the NUMA effect.
SABER~\cite{SABER} focuses on efficiently realizing computing power from both CPU and GPUs. 
Streambox~\cite{StreamBox} provides an efficient mechanism to handle out-of-order arrival event processing in a multi-core environment. 
Those solutions are complementary to ours and can be potentially integrated together to further improve DSPSs on shared-memory multicore architectures.

\emph{Execution plan optimization:}
Both operator placement and operator replication are widely investigated in the literature under different assumptions and optimization goals~\cite{4670119}.
In particular, many algorithms and mechanisms~\cite{Adaptive,T-storm,R-storm,Cardellini:2016:OOP:2933267.2933312,Cardellini:2017:OOR:3092819.3092823,Khandekar:2009:COS:1656980.1657002} are developed to allocate (i.e., schedule) operators of a job into physical resources (e.g., compute node) in order to achieve a certain optimization goal, such as maximizing throughput, minimizing latency or minimizing resource consumption, etc. Due to space limitation, we discuss them in Appendix~\ref{sec:more}.
Based on similar ideas from prior works, we implement algorithms including FF that greedily minimizes communication and RR that tries to ensure resource balancing among CPU sockets. 
As our experiment demonstrates, both algorithms result in poor performance compared to our RLAS approach in most cases because they are often trapped in local optima.

\compact
\section{Conclusion}
\label{sec:conclusion}
We have introduced \system, a new data stream processing system  with a new streaming execution plan optimization paradigm, namely Relative-Location Aware Scheduling (RLAS). 
\system successfully scales stream computation towards hundred of cores under NUMA effect.
The experiments on eight-sockets machines confirm that \system significantly outperforms existing open-sourced DSPSs up to an order of magnitude. 

\section{acknowledgement}
The authors would like to thank the anonymous reviewers for their valuable comments.
This work is supported by a MoE AcRF Tier 2 grant (MOE2017-T2-1-122) and an NUS startup grant in Singapore.
Jiong He is supported by the National Research Foundation, Prime Ministers Office, Singapore under its Campus for Research Excellence and Technological Enterprise (CREATE) programme.
Chi Zhou's work is partially supported by the National Natural Science Foundation of China under Grant 61802260 and the Guangdong Natural Science Foundation under Grant 2018A030310440.

	\bibliographystyle{ACM-Reference-Format}
	\bibliography{sigproc}
\newcompact
\appendix
%

\section{Implementation details}
\label{sec:impl_detail}
BriskStream shares many similarities to existing DSPSs including pipelined processing and operator replication designs. 
To avoid reinventing the wheel, we reuse many components found in existing DSPSs such as Storm, Heron and Flink, notably including \emph{API design, application topology compiler, pipelined execution engine with communication queue and back-pressure mechanism}.
In contrast, BrickStream embrace various designs that are suitable for shared-memory multicore architectures. For example, Heron has an operator-per-process execution environment, where each operator in an application is launched as a dedicated JVM process. In contrast, an application in BriskStream is launched in a JVM process, and operators are launched as Java threads inside the same JVM process, which avoids cross-process communication and allows the pass-by-reference message passing mechanism. 
Specifically, tuples produced by an operator are stored locally, and pointers as reference to tuple are inserted into a communication queue. Together with the jumbo tuple design, reference passing delay is minimized and becomes negligible.

\begin{figure*}
    \centering
   	\begin{minipage}{0.33\textwidth}
    \includegraphics*[width=\textwidth]{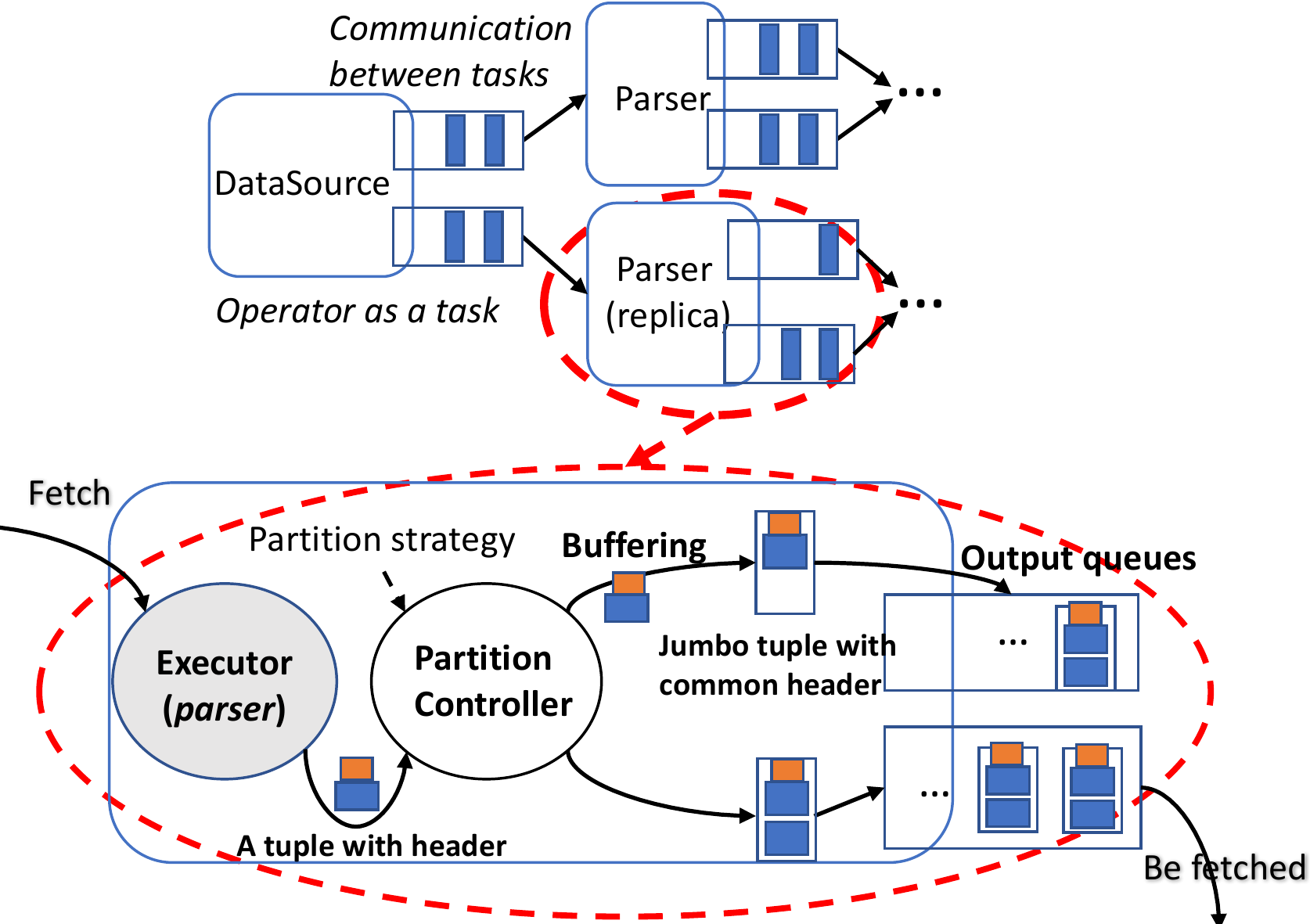}
    \caption{An example job in BriskStream.}  \label{fig:overview}   
     \end{minipage}   
    \begin{minipage}{0.36\textwidth}
        \subfloat[Fraud Detection (FD)]{%
            \includegraphics*[width=0.45\textwidth]{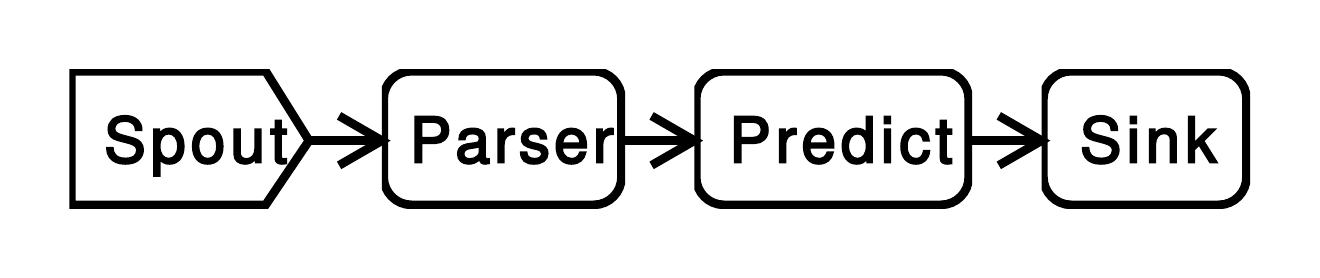}
        }
        \subfloat[Spike Detection (SD)]{%
            \includegraphics*[width=0.55\textwidth]{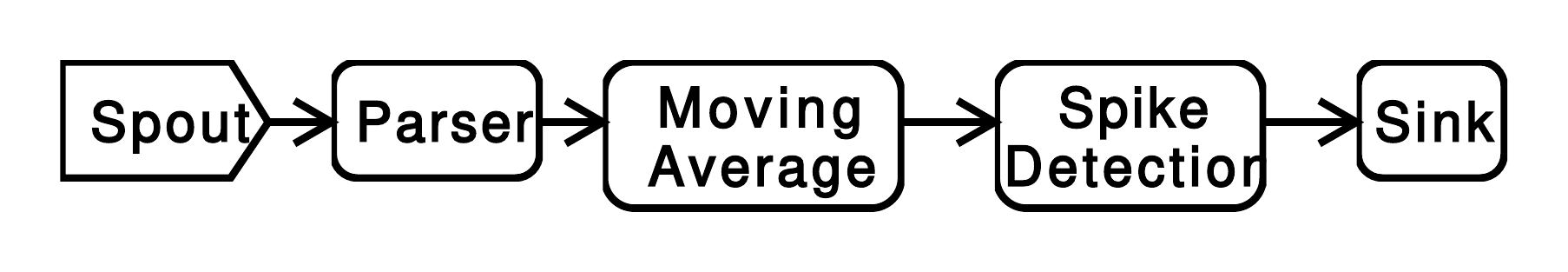}
        }
        
        \subfloat[Linear Road (LR)]{%
            \includegraphics*[width=\textwidth]{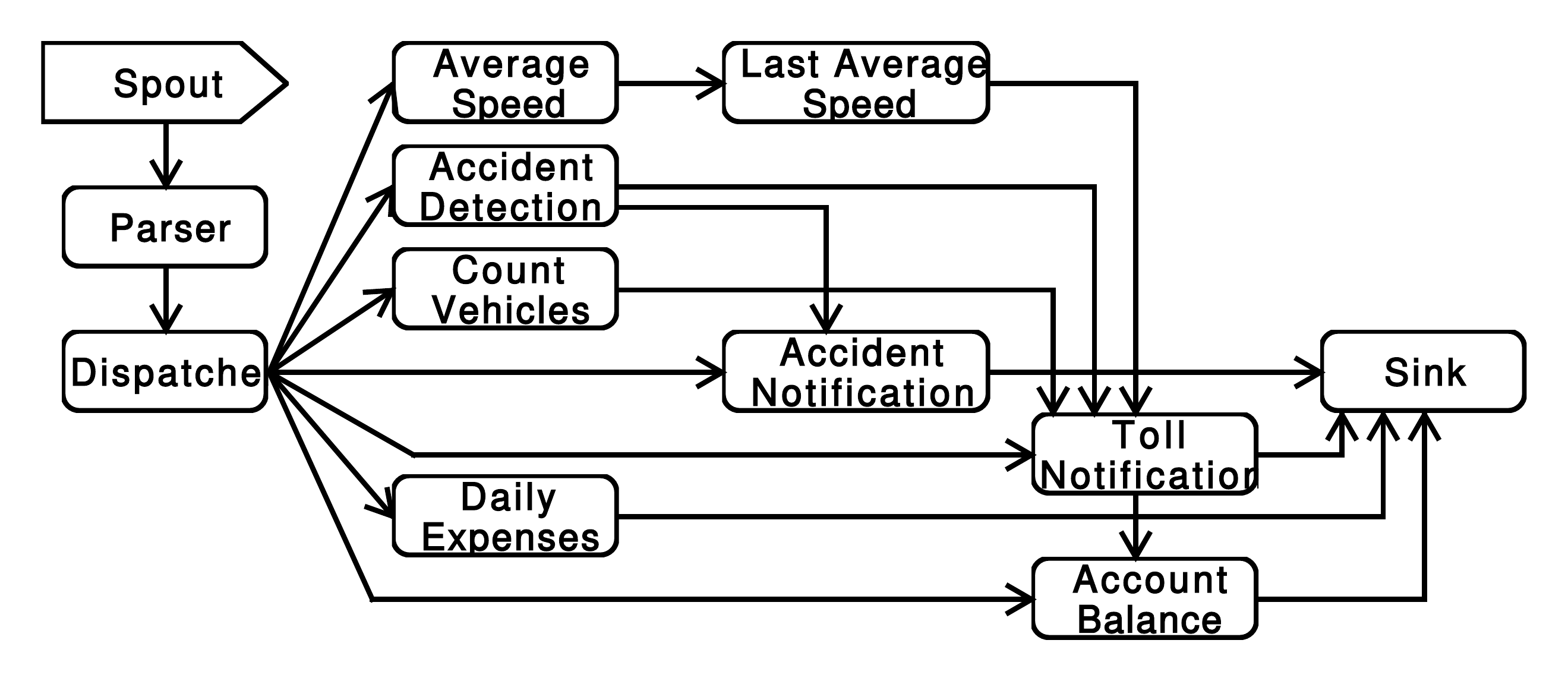}\label{fig:lr}
        }
        \caption{Topologies of other three applications in our benchmark.}\label{fig:topology}  
    \end{minipage} 
    \begin{minipage}{0.3\textwidth}
        \centering
        \captionof{table}{Operator selectivity of LR}
        \label{fig:selectivity}
        \includegraphics[width=\textwidth]{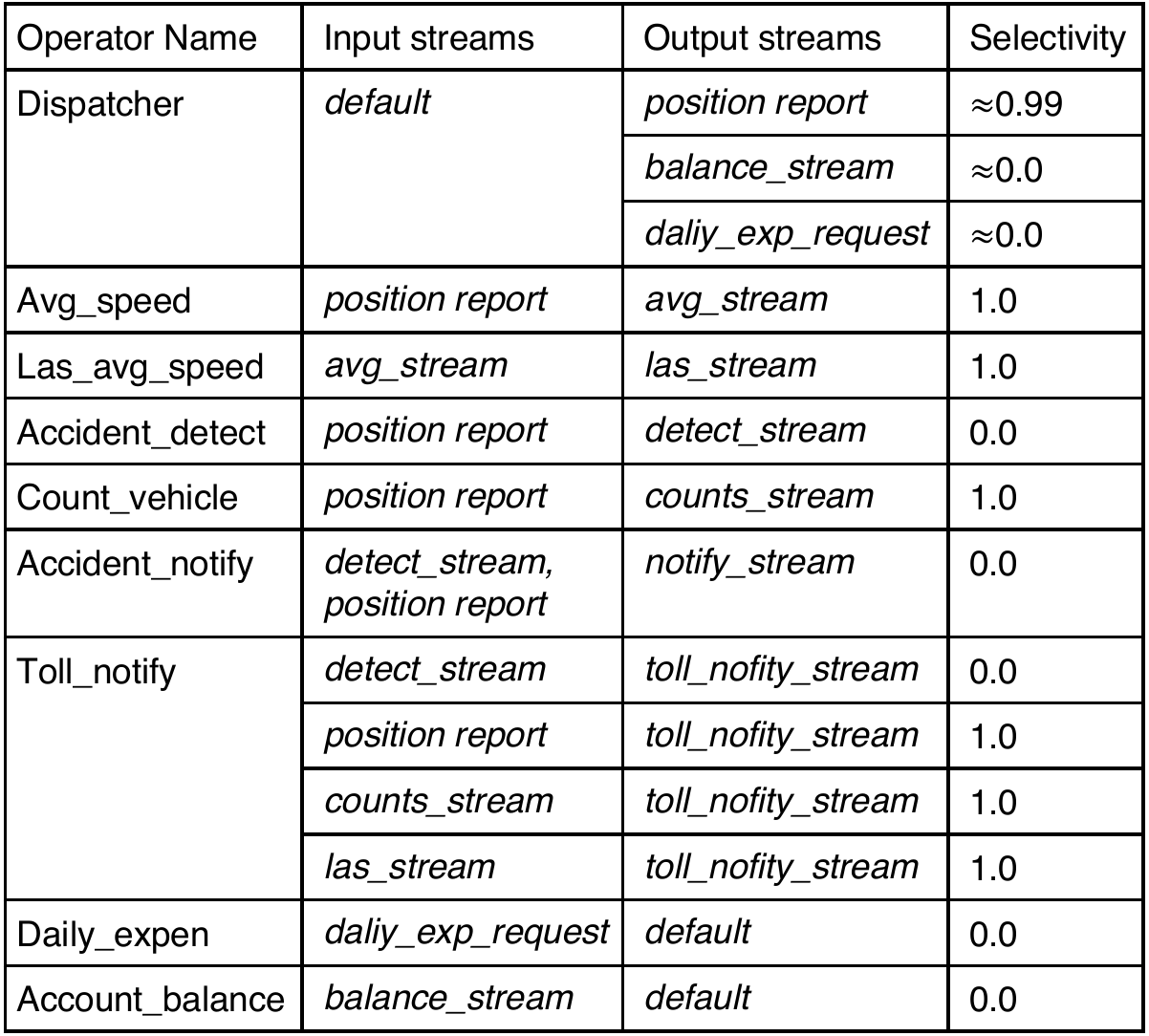}         
    \end{minipage}  
\end{figure*}

Figure~\ref{fig:overview} presents an example job (WC) of BriskStream. 
Each operator (or the replica) of the application is mapped to one~\emph{task}. 
The task is the basic processing unit in BrickStream (i.e., executed by a Java thread), which consists of an \emph{executor} and a \emph{partition controller}. 
The core logic for each {executor} is provided by the corresponding operator of the application. 
Executor operates by taking a tuple from the output queues of its producers and invokes the core logic on the obtained input tuple. 
After the function execution finishes, it dispatches zero or more tuples by sending them to its {partition controller}.
The partition controller decides in which output queue a tuple should be enqueued according to application specified partition strategies such as shuffle partitioning. 
Furthermore, each task maintains output buffers for each of its consumers, where jumbo tuples are formed accordingly. 


\newcompact
\section{Application Settings}
\label{subsec:application}
In this section, we discuss more settings of the testing applications in our experiment.
We have shown the topology of WC in Figure~\ref{fig:example}. Figure~\ref{fig:topology} shows the topology of the other three applications. More details about the specification about them can be found in the previous paper~\cite{profile}.

%

The selectivity is affected by both input workloads and application logic. 
Parser and Sink have a selectivity of one in all applications. 
Splitter has a output selectivity of ten in WC. That is, each input sentence contains 10 words. Counter has an output selectivity of one, thus it emits the counting results of each input word to Sink.
Operators have an output selectivity of one in both FD and SD. 
That is, we configure that a signal is passed to Sink in both predictor operator of FD and Spike detection operator of SD regardless of whether detection is triggered for an input tuple.
Operators may contain multiple output streams in LR. 
If an operator has only one output stream, we denote its stream as the \emph{default} stream.
We show the selectivity of each output stream of them of LR in Table~\ref{fig:selectivity}. 


\newcompact
\section{Algorithm Details}
\label{subsec:algo_impl}
In this section, we first present the detailed algorithm implementations including operator replication optimization (shown in Algorithm~\ref{alg:algo2}) and operator placement (shown in Algorithm~\ref{alg:algo}).
After that, we discuss observations made in applying algorithms in optimizing our workload and their runtime (Appendix~\ref{subsec:discuss}). 
We further elaborate how our optimization paradigm can be extended with other optimization techniques (Appendix~\ref{subsec:extention}).

Algorithm~\ref{alg:algo2} illustrates our scaling algorithm based on topological sorting. 
Initially, we set replication level of each operator to be one (Lines 1$\sim$2).
The algorithm proceeds with this and it optimizes operator placement with Algorithm \ref{alg:algo} (Line 6). 
Then, it stores the current plan if it ends up with better performance (Lines 7$\sim$8). 
At Lines 11$\sim$19, we iterate over all the sorted list from reversely topologically sorting on the execution graph in parallel (scaling from sink towards spout). 
At Line 15, it tries to increase the replication level of the identified bottleneck operator (i.e., this is identified during placement optimization). The size of increasing step depends on the ratio of over-supply, i.e., $\lceil \frac{r_i}{r_o} \rceil$.
It starts a new iteration to look for a better execution plan at Line 17.
The iteration loop ensures that we have gone through all the way of scaling the topology bottlenecks. 
We can set an upper limit on the total replication level (e.g., set to the total number of CPU cores) to terminate the procedure earlier. 
At Lines 9\&19, either the algorithm fails to find a plan satisfying resource constraint or hits the scaling upper limit will cause the searching to terminate.

\begin{algorithm}
\scriptsize
 \KwData{Execution Plan: $p$ \tcp{the current visiting plan}}
 \KwData{List of operators: $sortedLists$}
 \KwResult{Execution Plan: $opt$ \tcp{the solution plan}}
 $p.parallelism$ $\leftarrow$ set parallelism of all operators to be 1\;
 $p.graph$ $\leftarrow$ creates execution graph according to $p.parallelism$\;
 $opt.R$ $\leftarrow$ 0\;
 \KwRet Searching($p$);
 \caption{Topologically sorted iterative scaling} \label{alg:algo2}
  \SetKwFunction{FMain}{Searching}
  \SetKwProg{Fn}{Function}{:}{}
  \Fn{\FMain{$p$}}{
  $p.placement$ $\leftarrow$ placement optimization of $p.graph$\;     
  \If{$p.R>opt.R$}{$opt$ $\leftarrow$ $p$\ \tcp{update the solution plan}}
  \If{failed to find valid placement satisfying resource constraint}{\KwRet opt\;}
      $sortedLists$ $\leftarrow$ reverse TopologicalSort ($p.graph$)\tcp{scale start from sink}
      \ForEach{$list$ $\in$ $sortedLists$}{
          \ForEach{Operator $o$ $\in$ list}{
              \If{$o$ is bottleneck}{
                   $p.parallelism$ $\leftarrow$ try to increase the replication level of $o$ by $\lceil \frac{r_i}{r_o} \rceil$\;
                   \eIf{suceessfully increased $p.parallelism$}{\KwRet Searching($p$) \tcp{start another iteration}}{\KwRet opt}              
               }
          }
      }    
      \KwRet opt\;
  }
\end{algorithm}

Algorithm~\ref{alg:algo} illustrates our \emph{Branch and Bound based Placement} algorithm. 
Initially, no solution node has been found so far and we initialize a root node with a plan collocating all operators (Lines 1$\sim$5).
At Lines 7$\sim$14, the algorithm explores the current node. 
If it has better bounding value than the current solution, we update the solution node (Lines 10$\sim$11) if it is valid (i.e., all operators are allocated),
or we need to further explore it (Line 13). Otherwise, we prune it at Line 14 (this also effectively prunes all of its children nodes).
The branching function (Lines 15$\sim$32) illustrates how the searching process branches and generates children nodes to explore. 
For each collocation decision in the current node (Line 16), we apply the \emph{best fit heuristic} (Lines 17$\sim$23) and one new node is created. 
Otherwise, at Lines 25$\sim$27, we have to create new nodes for each possible way of placing the two operators (i.e., up to $\binom{m}{1}*\binom{2}{1}$). 
At Line 28$\sim$ 31, we update the number of valid operators and bounding value of each newly created nodes in parallel. 
Finally, the newly created children nodes are pushed back to the stack.

\begin{algorithm}[t]
\scriptsize
 \KwData{Stack $stack$ \tcp{stors all live nodes}}
 \KwData{Node $solution$ \tcp{stores the best plan found so far}}
 \KwData{Node $e$\tcp{the current visiting node}}
 \KwResult{Placement plan of $solution$ node}
 \tcp{Initilization}
 $solution.R$ $\leftarrow$ $0$ \tcp{No solution yet} 
 $e.decisions$  $\leftarrow$ a list contains all possible collocation decisions\;
 $e.plan$ $\leftarrow$ all operators are collocated into the same CPU socket\;
 $e.R$ $\leftarrow$ BoundingFunction($e.plan$)\;
 $e.validOperators$ $\leftarrow$ 0\;
 Push($stack$,$e$)\; 
 \While{$\neg$IsEmpty($stack$)\tcp{Branch and Bound process}}{ 
     $e$ $\leftarrow$ Pop($stack$)\;
  \eIf{$e.R > solution.R$}
   {
    \eIf{$e.validOperators$ == totalOperators}{$solution$ $\leftarrow$ $e$\;}{Branching(e)\;}
   }{\tcp{the current node has worse bounded value than solution, and can be safely pruned.}}
 }
 \caption{B\&B based placement optimization} \label{alg:algo}
  \SetKwFunction{FMain}{Branching}
  \SetKwProg{Fn}{Function}{:}{}
  \Fn{\FMain{$e$}}{
   \KwData {Node[] $children$}
   \ForEach{pair of $O_s$ and $O_c$ in $e.decisions$}{
       \eIf{all predecessors of them are already allocated except $O_s$ to $O_c$}{
           $\#newAllocate$ $\leftarrow$ 2\;
           \eIf{they can be collocated into one socket}{
               create a Node $n$ with a plan collocating them to one socket\;  
           }{
               create a Node $n$ with a plan separately allocating them to two sockets\; 
           }
           add $n$ to $children$\;   
       }
       {
           $\#newAllocate$ $\leftarrow$ 1\;
           \ForEach{valid way of placing $O_s$ and $O_c$}
           {create a new Node and add it to $children$;}    
       }
   }
       \ForEach{Node $c$ $\in$ $children$\tcp{update in parallel}}{
           $c.validOperators$ $\leftarrow$ e.validOperators + $\#newAllocate$\;
           $c.R$ $\leftarrow$ BoundingFunction($c.plan$)\;
       }
       PushAll($stack$, $children$)\;   
  }
\end{algorithm}

\section{Discussions on Optimization Process}
\label{subsec:discuss}
We have made some interesting observations in optimizing our workload. 
\emph{First}, placement algorithm (Algorithm~\ref{alg:algo}) start with no initial solution (i.e., the $solution.value$ is $0$ initially at Line 9) by default, and we have tried to use a simple first-fit (FF) placement algorithm to determine an initial solution node to potentially speed up the searching process. 
In some cases, it accelerates the searching process by earlier pruning and makes the algorithm converges faster, but in other cases, the overhead of running the FF algorithm offsets the gain.
\emph{Second}, the placement algorithm may fail to find any valid plan as it is not able to allocate one or more operators due to resource constraints, which causes scaling algorithm to terminate.
It is interesting to note that this \emph{may not} indicate the saturation of the underlying resources but the operator itself is too coarse-grained. 
The scaling algorithm can, instead of termination (in Line 10 Algorithm~\ref{alg:algo2}), try to further increase the replication level of operator that ``failed-to-allocate''. After that, workloads are essentially further partitioned among more replicas and the placement algorithm may be able to find a valid plan. This procedure, however, introduces more complexity to the algorithm.
    
\textbf{Optimization runtime.}
The placement optimization problem is difficult to solve as the solution space increases rapidly with increased replication level configurations. 
Besides the three proposed heuristics, we also apply a list of optimization techniques to further increase the searching efficiency including 1) memorization in evaluating performance model under a given execution plan (e.g., an operator should behave the same if its relative placement with all of its producers are the same in different plans), 2) instead of starting from scaling with replication set to one for all operators, we can start from a reasonable large DAG configuration to reduce the number of scaling iteration and 3) the algorithm is highly optimized for higher concurrency (e.g., concurrently generate branching children nodes).
Overall, the placement algorithm needs less than $5$ seconds to optimize placement for a large DAG, and the entire optimization usually takes less than $30$ seconds, which is acceptable, given the size of the problem and the fact that the generated plan can be used for the whole lifetime of the application. 
As the streaming application usually runs for a long time, the overhead of generating a plan is not included in our measurement.

\textbf{Extension with other optimization techniques.}
\label{subsec:extention}
A number of optimization techniques are available in the literature~\cite{optimization,Deployment}. 
Many of them can be potentially applied to further improve the performance of BriskStream. 
Taking operator fusion as an example, which trades communication cost against pipeline parallelism and is in particular helpful if operators share little in common computing resource. Our performance model is general enough such that it can be extended to capture other optimization techniques. 


\newcompact

\section{More Related Work}
\label{sec:more}
Aniello et al.~\cite{Adaptive} propose two schedulers for Storm. 
The first scheduler is used in an offline manner prior to executing the topology and the second scheduler is used in an online fashion to reschedule after a topology has been running for a duration. 
Similarly, T-Storm~\cite{T-storm} dynamically assigns/reassigns operators according to run-time statistics in order to minimize inter-node and inter-process traffic while ensuring load balance. 
R-Storm~\cite{R-storm} focuses on resource awareness operator placement, which tries to improve the performance of Storm by assigning operators according to their resource demand and the resource availability of computing nodes. 
Cardellini et al.~\cite{Cardellini:2016:OOP:2933267.2933312,Cardellini:2017:OOR:3092819.3092823} propose a general mathematical formulation of the problem of optimizing operator placement for distributed data stream processing. 
Those approaches may lead to a suboptimal performance in the NUMA environment that we are target at. 
This is because factors including output rate, amount of communication as well as resource consumption of an operator may change in different execution plans due to the NUMA effect and can therefore \emph{mislead} existing approaches that treat them as predefined constants. 

\end{document}